\documentstyle[galley,epsfig]{mn2e}
\title{Chemical  analysis of  CH stars - III: atmospheric parameters and elemental abundances}
\title[Chemical  analysis of  CH stars - III: atmospheric parameters and elemental abundances\thanks ]
{Chemical analysis of  CH stars - III: atmospheric parameters and elemental abundances}
\author[Meenakshi Purandardas  et al.]{Meenakshi Purandardas$^{1,2}$, Aruna Goswami$^{1}$ \thanks{E-mail: aruna@iiap.res.in}, Partha Pratim Goswami$^{1}$, 
\newauthor Shejeelammal J$^{1}$, Thomas Masseron$^{3,4}$  \\
    $^{1}$Indian Institute of Astrophysics, Koramangala, Bangalore 560034,
    India;  aruna@iiap.res.in\\ 
$^{2}$ Department of physics, Bangalore university, Jnana Bharathi Campus, Karntaka 560056, India\\
$^{3}$ Instituto de Astrofisica de Canarias, E-38205 La Laguna, Tenerife, Spain \\
$^{4}$ Departamento de Astrofisica, Universidad de La Laguna, E-38206 La Laguna, Tenerife, Spain. 
}

\begin{document}

\date{ Accepted ---;  Received ---;  in original
form --- \large \bf }

\pagerange{\pageref{firstpage}--\pageref{lastpage}} \pubyear{2014}

\maketitle


\begin{abstract}
Elemental abundances of CH stars can provide observational constraints
for theoretical studies on the nucleosynthesis and evolution of 
low- and intermediate-mass stars. However, available abundance data 
in literature are quite scanty. In our pursuit to generate a 
homogeneous database  of elemental abundances of  CH stars  we have performed 
a detailed chemical abundance study for a sample of 12 potential 
CH star candidates based on high resolution spectroscopy. We present 
first time abundance analysis  for the objects HE0308$-$1612,
 CD$-$281082, HD30443, and HD87853. 
For the other objects, although  limited information is available,  
detailed chemical composition studies are  missing. Our 
analysis shows CD$-$281082 to be a very metal-poor object with 
[Fe/H]=$-$2.45 and  enriched  in carbon with [C/Fe] = 2.19. With 
a ratio of [Ba/Eu]${\sim}$0.02  the star satisfies the 
classification criteria of a CEMP-r/s star. The objects CD$-$382151  
with [Fe/H]=$-$2.03 and HD30443 with [Fe/H]${\sim}$$-$1.68
are found to show the characteristic properties  of CH stars. 
HE0308$-$1612 and HD87853 are found to be moderately
metal-poor with [Fe/H]${\sim}$$-$0.73; while HE0308$-$1612 is moderately
enhanced with carbon ([C/Fe]${\sim}$0.78) and shows the spectral
properties of CH stars,  the abundance of carbon
could not be estimated for HD87853.  
Among the two  moderately metal-poor objects, HD87080 ([Fe/H]=$-$0.48) 
shows near solar carbon abundance,  and  HD176021 ([Fe/H]=$-$0.63 ) 
is mildly enhanced in carbon with [C/Fe]=0.52. HD176021
along with HD202020 a known binary,   exhibit the 
characteristic properties of CH stars as far as the  heavy element 
abundances are concerned. Five objects in our sample show spectral 
properties that are normally seen in barium stars. 
\end{abstract}


\section{Introduction}

In a recent study by Cristallo et al. (2016), a homogeneous set of
abundance data from CH stars (Goswami et al. 2006; Goswami, Karinkuzhi \& Shantikumar 2010; 
Karinkuzhi \& Goswami 2014, 2015; Goswami, Aoki \& Karinkuzhi 2016) are used in an attempt to 
constrain the physics and the nucleosynthesis occurring in low-mass 
AGB stars. The scope of enhancing such studies based on a larger 
homogeneous  abundance data was  highlighted. Particularly, the  data 
corresponding to the lower metallicity  regime were found to be scanty.
In our efforts to provide tighter observational constraints  we have 
undertaken to conduct chemical composition study for a larger 
sample selected  from the CH star catalogue of Bartkevicius (1996), and
our previous work (Goswami 2005, Goswami et al. 2007, 2010).
In this work we report results  for twelve objects.  As will be 
shown later in detail,  we have found  two objects  to be very 
metal-poor stars with [Fe/H] $<$ $-$2 and  enhanced in carbon 
with [C/Fe] $>$ 1. Five  objects  exhibit 
characteristic properties of CH stars, and the other five
show spectral properties that are normally seen in barium stars.

CH stars characterized by iron-deficiency 
 and carbon enhancement (with $-$2 $\le$ [Fe/H] $\le$ $-1$, 
[C/Fe] ${\ge}$ 0.7,  [Ba/Fe] ${\ge}$ 1.0) are a distinct group of 
stars (Keenan 1942) that are long been used as halo tracers. 
CEMP-s stars as defined in Beers \& Christlieb (2005), classified 
as stars  
with [C/Fe] ${\textgreater}$ 1 and   [Fe/H] ${\textless}$ $-$2
are more metal-poor counterparts of  CH stars and believed to 
be extrinsically  enriched with carbon and  s-process elements. 
From observational evidences,  CH stars are known to be binary 
systems (McClure 1983, 1984, McClure \& Woodsworth 1990), with 
a now invisible white dwarf companion.  Until recently, the 
CEMP-s stars with enhanced abundances of heavy  s-process 
elements and carbon are also thought  to be  binary systems 
(Lucatello et al. 2005,  Bisterzo et al. 2011, Starkenburg 
et al. 2014, Jorissen et al. 2016) and  inferred to have 
 a similar origin as that of the CH stars. However, based on a 
long-term campaign of precision radial-velocity monitoring of 
a sample of 22  CEMP-s and CEMP-r/s stars,  Hansen et al. (2016) 
reported   18 of them  (82 percent $\pm$ 26 percent) to be bonafide binary 
systems; the  rest  are either single or objects with  extremely 
long period. If single, with carbon and barium enhancement 
these objects draw special attention for an intrinsic origin. 
Among many  scenarios, the ones involving  massive stars producing
barium and other heavy elements  through a weak s-process mechanism 
are  suggested by many authors  as possible mechanism(s) 
(Frischknecht et al. 2016, and  references therein).
These suggestions however require detailed investigation and 
larger samples  are clearly desirable. The  enhanced abundances 
of heavy elements observed in the (secondary) star of the 
CH binary star systems that are in the giant phase of evolution 
are attributed to a now invisible white dwarf companion (primary) 
which has  evolved   through the Asymptotic Giant Branch (AGB) phase. 
During the AGB phase  
the heavy elements were synthesized  and the synthesized 
 materials are  transferred through mass transfer mechanisms 
to the secondary. Although the mass transfer mechanism(s) are 
not well understood  despite several dedicated studies 
(e.g., Abate et al. 2013, 2015a,b, and references therein)
the surface chemical composition of the secondaries can be 
used as  important diagnostics for s-process nucleosynthesis  
in the primary. 

\par The 12 CH star candidates analysed in this study 
cover a wide range in metallicity from near solar to [Fe/H] = $-$2.4. 
While 
CD$-28$ 1082, and  CD$-38$ 2151 are found to be very metal-poor with
[Fe/H] $<$ $-$2.0,  HD~176021 and HD~87080 are found to be mildly 
metal-poor  with metallicities $-$0.63 and $-$0.48 respectively. 
We have  investigated 
anew the elemental abundances of two objects  (HD ~87080, 
HD~188985) previously studied by Allen \& Barbuy (2006a) 
and another two objects (HD~29370, HD~123701) previously studied 
by de Castro et al. (2016)  that  allowed us to make a direct 
comparison with ours  and verify  our estimates. 
HD 87080 was also  analysed by Pereira \& Junqueira (2003);
they  found that iron group, $\alpha$-elements, manganese, 
as well as sodium and aluminium of  HD 87080 follow the abundance
pattern of the Galactic disk stars.  Analysis of North, Berthet \& Lanz (1994a,b) 
have  confirmed  HD 188985  to be a barium dwarf.  We have also presented 
abundances for C, N, and O for  these objects that were  not reported in 
their previous studies.
\vskip 0.2cm
\noindent
In section 2,  we briefly describe the source of the spectra used 
in the present study and present estimates of radial velocities.  
Section 3 gives the estimates of photometric temperature. Section 4 
describes the methodology used for determination of atmospheric 
parameters and  estimates of the  elemental abundances. This section
also provides a discussion on the estimates of stellar masses and age of the
programme stars. A  comparison  between the estimates   obtained 
from spectra using
  VBT echelle and    FEROS  is  presented
in section 5. In section 6  we discuss  the abundance analysis.
In section 7, we discuss the kinematic analysis. 
stars are presented in section 8 and  conclusions are drawn  in section 9. 

\section{Observations and data reduction}
For the 
objects  HD~29370, CD$-$38~2151, HD~50264, HD~87080,  HD~123701, and 
HD~188985, high resolution (${\lambda}/{\delta\lambda}$ ${\sim}$ 72000) 
spectra  were  obtained during 2018 March - May using
 the high resolution fiber fed Echelle spectrometer attached to the
2.34 m Vainu Bappu Telescope (VBT) at the Vainu Bappu Observatory (VBO),
 Kavalur. The wavelength coverage of these spectra spans from 
 4100 to 9350 {\rm \AA} with  gaps between orders.
The data-reduction is done following the standard procedures
using  IRAF\footnote{IRAF is distributed by the National
Optical Astronomical Observatories, which is operated by the Association
for Universities for Research in Astronomy, Inc., under contract to the
National Science Foundation} spectroscopic data reduction  package.
A few  high-resolution spectra  obtained with 
the FEROS (Fiber-fed Extended Range Optical Spectrograph) of the 
1.52 m telescope of ESO at La Silla, Chile are also used in  the 
present study.  The wavelength coverage of 
FEROS spectra spans  from 3500 - 9000 {\rm \AA}
and the  spectral resolution  is ${\sim}$ 48000.
For the objects HD~30443 and HE~0308$-$1612, the high resolution 
($\lambda/\delta\lambda \sim 60000 $) spectra were obtained during
2017 November and 2018 January using the high-resolution fiber 
fed  Hanle Echelle SPectrograph (HESP) attached to the 2 m Himalayan 
Chandra Telescope (HCT) at the Indian Astronomical Observatory, Hanle. 
The wavelength coverage of these spectra spans from 3530-9970 {\rm \AA}. 
Table 1 lists the basic data for the programme stars and details of 
the source of spectra. A few sample spectra are shown in Figure 1.

\begin{figure}
\centering
\includegraphics[width=\columnwidth]{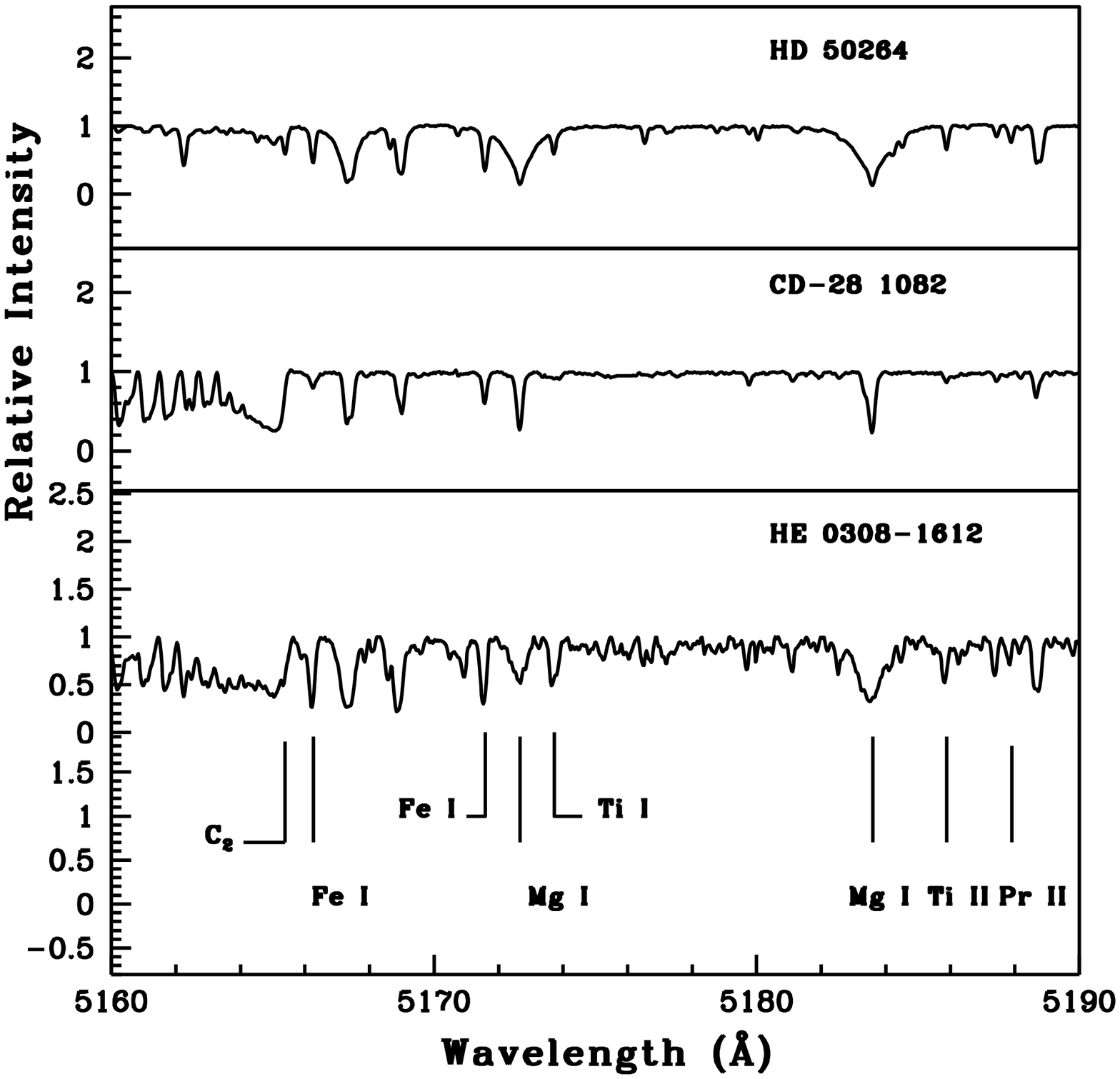}
\caption{ Examples of sample spectra of the programme stars  in the  
wavelength region 5160 to 5190 {\bf  {\rm \AA}}. In the upper panel the 
spectrum of a barium star, in the middle panel the spectrum of a CEMP-r/s 
star and in the 
bottom panel the spectrum of a CH star are shown. }
\end{figure}

{\footnotesize
\begin{table*}
{\bf Table 1: Basic data for the programme stars}\\ 
\begin{tabular}{lcccccccccc}
\hline
Star      &RA$(2000)$       &Dec.$(2000)$    &B       &V       &J        &H        &K  &Exposure   &Date of obs.& Source  \\
          &                 &                &        &        &         &         &   & (seconds)           &            & of spectrum\\
\hline

HE 0308$-$1612  & 03 10 27.07 &$-$16 00 40.1 & 12.7   & --     & 10.03   & 9.48    & 9.33  &2700(4) & 24-01-2018 &HESP \\
CD$-$28 1082  &03 16 26.43  &$-28$ 10 48.02  &11.55   &10.41   &8.89     &8.44     &8.30   &3600 &05-01-2000& FEROS\\
HD 29370      &04 35 40.04  &$-42$ 00 05.58  &10.43   &9.33    &7.57     &7.08     &6.97   &2700 &04-01-2000& FEROS \\
              &             &                &        &        &         &         &       &2700 (2) &07-03-2018& VBT Echelle \\
HD 30443      &04 49 16.02  &+35 00 6.49     &11.2    &8.82    &4.62     &3.28     &2.83   &2400(2)  &07-11-2017 & HESP \\
CD$-$38 2151 &05 40 01.94  &$-38$ 49 25.89  &10.75   &9.27    &7.16     &6.56     &6.39   &3600 &06-01-2000& FEROS\\
              &             &                &        &        &         &         &       &2700 (2) &06-03-2018& VBT Echelle \\
HD 50264      &06 51 12.49  &$-29$ 34 32.35  &9.63    &9.05    &7.91     &7.61     &7.57   &1200 &10-11-1999& FEROS\\
              &             &                &        &        &         &         &       &2700(2) &6-03-2018& VBT Echelle \\
HD 87080      &10 02 00.85  &$-33$ 41 06.50  &10.15   &9.37    &8.00     &7.68     &7.56   &2100 &05-01-2000& FEROS\\
              &             &                &        &        &         &         &       &2700(3) &09-04-2018& VBT Echelle \\
HD~87853      & 10 07 58.78 & +08 30 37.24   & 10.12  & 9.66   & 8.81    & 8.57    & 8.55  & 1800   & 06-01-2000  & FEROS \\
HD 123701     &14 09 55.10  &$-30$ 26 33.91  &9.54    &8.51    &6.78     &6.30     &6.21   &1200 &05-01-2000& FEROS\\
              &             &                &        &        &         &         &       &2700(3)  &06-03-2018& VBT Echelle \\
HD 176021     &19 02 37.30  &$-64$ 55 31.30  &8.20    &7.58    &6.47     &6.20     &6.11   &300  &14-07-2000& FEROS\\
HD 188985     &19 59 58.29  &$-48$ 58 31.05  &9.07    &8.56    &7.54     &7.32     &7.23   &600  &14-07-2000& FEROS\\
              &             &                &        &        &         &         &       &2700(1)  &09-04-2018& VBT Echelle \\
HD 202020     &21 13 39.47  &$-09$ 37 41.73  &9.90    &9.37    &8.15     &7.89     &7.82   &1200 &14-07-2000& FEROS\\  

\hline

\end{tabular}

\textit{Note.}The numbers in the parenthesis with exposures indicate the number of 
frames taken.\\
PIs of the original observing proposals are V. Hill for FEROS 1999 spectra,
F. Primas for FEROS 2000 spectra, and A.Goswami for VBT Echelle and 
HESP spectra.
\end{table*}
}

\section{Photometric temperatures}
The photometric temperatures of our programme stars are estimated using the 
calibration equations of Alonso, Arribas \& Martinez-Roger(1999, 2001) and following the
detailed procedures as given in our earlier papers (Goswami et al. 2006, 2016). 
 As our programme  star sample consists of both main-sequence stars and 
giants, we have made use of the corresponding calibration equations 
separately for each group.  J, H, K magnitudes of the objects required 
for temperature estimates are taken from SIMBAD database which came 
from 2MASS (Cutri et al. 2003). We have used the estimated photometric 
temperatures  as  the initial guess for the estimation of spectroscopic 
temperature of  the objects. The estimated photometric temperatures are 
presented in Table 2. 

{\footnotesize
\begin{table*}
{\bf Table 2 : Temperatures from  photometry }\\
\tiny
\begin{tabular}{llllllllllll}
\hline                      
Star name & $T_eff$   &$T_eff $& $T_eff$ & $T_eff$ & $T_eff$ & $T_eff$ & $T_eff$ & $T_eff$& $T_eff$ & $T_eff$ & Spectroscopic\\
          &           &($-0.05)$& ($-0.5$)& ($-1.5$)& ($-0.05$)& ($-0.5$)& ($-1.5$) & ($-0.05$) & ($-0.5$)& ($-1.5)$ &estimates  \\
          & (J$-$K)   &(J$-$H) & (J$-$H) & (J$-$H) & (V$-$K) & (V$-$K) & (V$-$K) &(B$-$V)  & (B$-$V) & (B$-$V) &  \\
\hline 
HE 0308-1612 & 4420.26 &4413.24 &4428.17 &4461.72 &   -    &  -     &   -    &  -     &  -     &   -    & 4600 \\
CD$-28$ 1082 & 4862.67 &5044.69 &5091.08 &5119.74 &4956.30 &4941.81 &4928.88 &4534.58 &4452.10 &4353.37 &5200\\
HD 29370     & 4830.66 &4874.24 &4916.69 &4941.53 &4703.69 &4685.78 &4663.45 &4604.81 &4515.99 &4406.40 &4970 \\
HD 30443     & -       &  -     &  -     &  -     &  -     &  -     &  -     &  -     & -      & -      & 4040 \\
CD$-38$ 2151 & 4358.16 &4465.73 &4499.41 &4545.89 &4292.39 &4275.88 &4255.08 &4015.86 &3975.85 & -      &4600\\
HD 50264     & 5964.48 &5890.11 &5903.72 &5934.18 &5778.41 &5773.23 &5780.91 &5892.20 &5735.39 &5509.55 &5900\\
HD 87080     & 5416.07 &5704.45 &5767.58 & -      &5295.93 &5286.78 &5288.41 &5257.63 &5133.97 &4934.15 &5600\\
HD 87853     & 6361.55 &6108.46 & 6121.51& 6150.71&6373.71 & 6377.72&6410.18 &6376.35 &6200.93 & 5953.78&6250 \\
HD 123701    &4928.34  &4915.63 &4959.03 &4984.78 &4762.04 &4744.87 &4724.61 &5691.86 &5568.66 &5378.63 &5340\\
HD 176021    & 5780.58 &6012.43 &6084.20 & -      &5739.06 &5738.20 &5762.19 &5691.86 &5568.66 &5378.63 &5900\\
HD 188985    &6132.50  &6461.37 &6473.32 &6500.05 &6018.47 &6016.72 &6033.72 &6165.46 &5998.25 &5760.42 &6250\\
HD 202020    & 5933.00 &6078.46 &6152.15 &-       &5628.51 &5625.44 &5643.49 &5978.92 &5856.51 &5674.51 &5960 \\
\hline      
\end{tabular}

\textit{Note.} The numbers in the parenthesis below $T_{eff}$ indicate the metallicity values at which the temperatures are 
calculated. Temperatures are given in Kelvin.

\end{table*}
}

\section{Spectroscopic stellar parameters}
Equivalent widths   of a set of clean and unblended Fe I and Fe II 
lines are used for the estimation of atmospheric parameters. The 
lines are  selected to have  the excitation potential  in the 
range 0.0-5.0 eV and  the equivalent widths in the range  
20-180 m\AA. The number of Fe II lines measured is small in 
number   as it is  difficult to get clean Fe II lines in the spectra. 
We have made use of Local Thermodynamic Equilibrium (LTE) analysis of the 
measured  equivalent widths using a recent version of MOOG of 
Sneden (1973). Model atmospheres used  are selected from Kurucz 
grid of model atmospheres with no convective overshooting 
(http://cfaku5.cfa.hardvard.edu/).  Solar abundances are adopted  
from Asplund, Grevesse \& Sauval (2009). 

A trend between the iron abundance derived from the Fe I lines and 
the excitation potential of these lines with a zero slope defines 
the effective temperature. The microturbulent velocity is taken 
to be that value for which the abundances  derived from the Fe I 
lines do not show any dependence on the reduced equivalent 
width (${W_{\lambda}/{\lambda}}$). The surface gravity log\,g is 
determined corresponding to the adopted values of the effective 
temperature and  microturbulent velocity for which Fe I and Fe II 
lines give near  equal abundance values.
The estimated  abundances from  Fe I and Fe II  
lines as  a function of excitation potential and equivalent 
widths, respectively, are shown in figures that are made available 
as on-line materials.  The derived atmospheric 
parameters are  listed in Table 3.

{\footnotesize
\begin{table*}
{\bf Table 3: Derived atmospheric parameters and radial velocities of 
the programme stars. }\\ 
\begin{tabular}{lcccccccc}
\hline
Star         &T$_{eff}$& log g &$\zeta$      & [Fe I/H]        &[Fe II/H]        & V$_{r}$     &  V$_{r}$     \\
             &    (K)  & cgs  &(km s$^{-1}$) &                 &                 & (km s$^{-1}$)& (km s$^{-1}$) \\
             &         &      &              &                 &                 & (our estimates)& (from literature)      \\
\hline
HE~0308$-$1612& 4600   & 1.7  & 1.42         &$-$0.72$\pm$0.19 & $-$0.73$\pm$0.15&  85.5$\pm$1.22 (HESP) &  76.41$\pm$0.51  \\
CD$-28$ 1082 & 5200    &1.90  &1.42          &$-2.46$$\pm$0.08 &$-2.44$$\pm$0.02 &  $-26.7$$\pm$0.3 (FEROS)&  --\\
HD 29370     & 4970    &2.40  &1.92          &$-0.15$$\pm$0.11 &$-0.13$$\pm$0.04 &  $-1.5$$\pm$0.03 (FEROS)& 16.0$\pm$2.6\\
HD 30443     & 4040    &2.05  &2.70             & $-1.68$$\pm$0.05   &$-1.69$$\pm$0.11 & 66.61$\pm$0.20 (HESP) & 63.17$\pm$0.22  \\ 
CD$-38$ 2151 & 4600    &0.90  &2.30          &$-2.03$$\pm$0.10 &$-2.03$          & 138.5$\pm$0.6 (FEROS)& 125  \\
             &         &      &              &                 &                 & 139.7$\pm$1.9 (VBT)  &                 \\
HD 50264     & 5900    &4.60  &0.62          &$-0.14$$\pm$0.09 &$-0.11$$\pm$0.07 &  65.8$\pm$0.4 (FEROS) & 56.9$\pm$0.4 \\
             &         &      &              &                 &                 &  67.8$\pm$2.6 (VBT)   &                   \\
HD 87080     & 5600    &3.74  &1.10          &$-0.47$$\pm$0.09 &$-0.49$$\pm$0.04 & 5.9$\pm$0.02 (FEROS) & $-0.5$$\pm$1.4\\
             &         &      &              &                 &                 & $-1.7$$\pm$0.5 (VBT) &                     \\
HD 87853     & 6250    &2.5   &1.63          &$-$0.72$\pm$0.09 &$-$0.74$\pm$0.09 & $-$3.35$\pm$0.05 (FEROS) &  --    \\
HD 123701    & 5340    &3.50  &1.44          &0.05$\pm$0.09    &0.07$\pm$0.09    & $-18.2$$\pm$0.1 (FEROS) & -- \\
             &         &      &              &                 &                 & $-24.8$$\pm$0.6 (VBT)   &           \\
HD 176021    & 5900    &3.95  &1.02          &$-0.62$$\pm$0.08 &$-0.65$$\pm$0.05 & 109.1$\pm$0.5 (FEROS) & 108.4$\pm$0.9\\
HD 188985    & 6250    &4.30  &0.60          &0.01$\pm$0.10    &0.01$\pm$0.09    & 6.7$\pm$0.03 (FEROS) & 8.2 \\  
HD 202020    & 5960    &4.38  &0.86          &0.04 $\pm$0.09   &0.08 $\pm$0.03   & $-18.6$$\pm$0.1 (FEROS) &$-24.5$$\pm$0.14\\
\hline
\end{tabular}

\end{table*}
}

We have estimated the mass of our programme stars from their 
postion in the HR diagram,
log$(L/L_{\odot})$ versus spectroscopic T$_{eff}$ plots (Figs 2-4). 
The visual magnitude V is taken from  SIMBAD  and the parallaxes are 
taken from \textit{Gaia} (Gaia collaboration 2016, 2018b, 
https://gea.esac.esa.int/archive/) whenever possible. 
The bolometric corrections  are determined separately for 
main-sequence and subgiants/giants using the empirical calibrations
of Alonso et al.(1999). Interstellar extinction (A$_{v}$) for 
objects with Galactic latitude  b $<$ 50 degree  are  calculated 
based on formulae  by Chen et al (1998). We made use of Girardi 
et al. (2000) database  (http://pleiadi.pd.astro.it/) of evolutionary 
tracks to estimate the mass of the stars.  For near solar objects 
evolutionary tracks corresponding to Z = 0.019 are chosen.  Surface gravity
log\,{g}  is  calculated from the estimated mass using the relation \\
log(g/g$_{\odot}$) = log(M/M$_{\odot}$) + 4log(T$_{eff}$/T$_{eff\odot}$) + 0.4(M$_{bol}$ $-$ M$_{bol\odot}$)\\
The adopted values for the Sun are log g$_{\odot}$ = 4.44, 
T$_{eff\odot}$ = 5770K, 
and M$_{bol\odot}$ = 4.75 mag (Yang et al. 2016). 
\par We have also determined the age of our programme stars from their 
locations 
in the HR diagram. We have made use of the Girardi et al. (2000) 
database of isochrones. We used isochrones corresponding to z = 0.0004 
 for CD$-28$ 1082, and z = 0.004 for HD 176021,  HD 87080 and HD~87853. 
For objects
with near solar metallicity, the isochrones corresponding to z = 0.019
are adopted. These are illustrated in Figures 5-7.
The age estimates are presented in Table 4.
\begin{figure}
\centering
\includegraphics[width=\columnwidth]{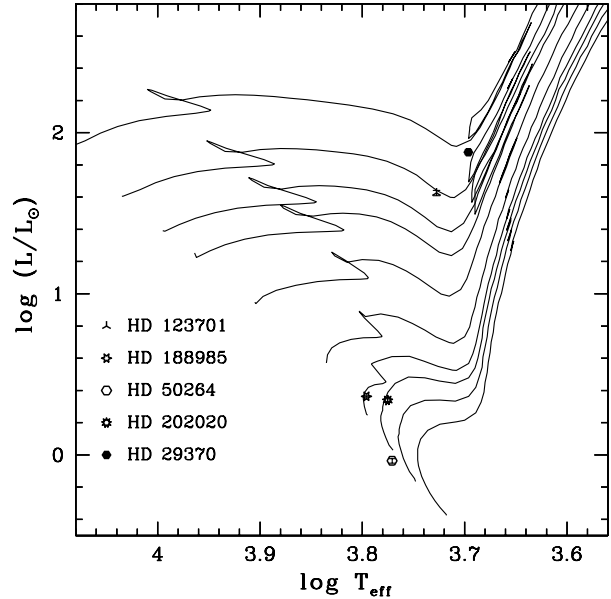}
\caption{The locations of HD 123701, HD 188985, HD 50264, HD 202020 
and HD 29370. The evolutionary tracks for 0.9, 1.0, 1.1, 1.2, 1.4, 
1.7, 2.0, 2.2, 2.5 and 3.0 M$_{\odot}$ are shown from bottom to top. }
\end{figure}

\begin{figure}
\centering
\includegraphics[width=\columnwidth]{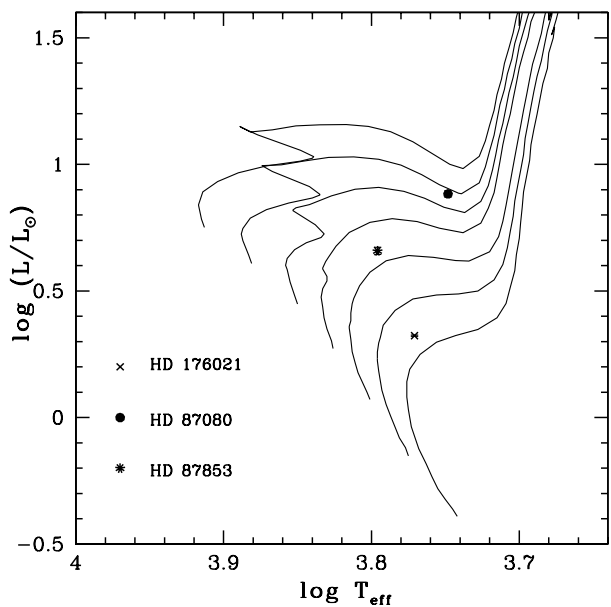}
\caption{The locations of  HD 87080, HD~87853 and  HD 176021. 
The evolutionary  tracks for 0.8, 0.9, 1.0, 1.1, 1.2, 1.3, 
and 1.4 M$_{\odot}$  are shown from  bottom to top. }
\end{figure}

\begin{figure}
\centering
\includegraphics[width=\columnwidth]{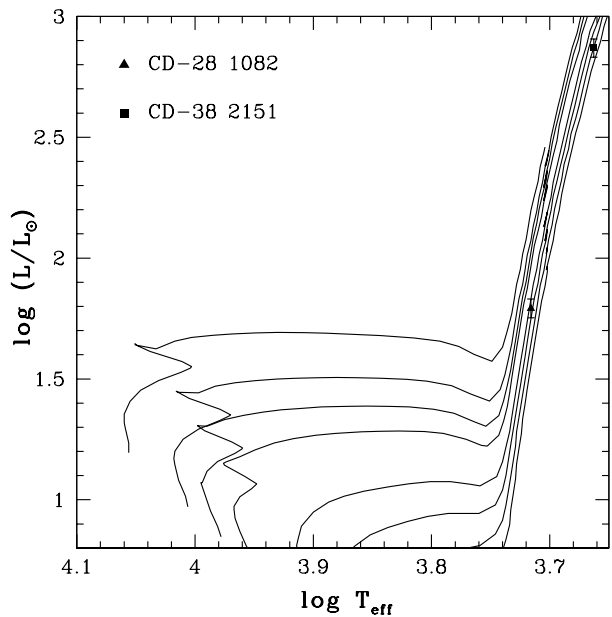}
\caption{The locations of CD$-28$ 1082 and CD$-38$ 2151. 
The evolutionary tracks for 0.8, 0.9, 1.0, 1.1, 1.3, 1.4, 
1.5 and 1.7  M$_{\odot}$ are shown from bottom to top. }
\end{figure}

{\footnotesize
\begin{table*}
{\bf Table 4: Estimates of  log\,{g} using parallax method }\\ 
\begin{tabular}{lccccccc}
\hline                       
 Star name    & Parallax        & $M_{bol}$    & log(L/L$_{\odot}$) & Mass(M$_{\odot}$) & log g & log g (spectroscopic)& Age \\
              & (mas)           &              &                    &                   & (cgs)  & (cgs)   & (Gyr)  \\
\hline
HE~0308$-$1612& 0.46$\pm$0.05   &  --          &  --                &  --             &  --           & 1.70  & -- \\
CD$-28$ 1082  & 1.09$\pm$0.05   & 0.27$\pm$0.09  & 1.79$\pm$0.04    & 1.05$\pm$0.05   & 2.52$\pm$0.03 & 1.9 & 7.9\\    
HD 29370      & 1.58$\pm$0.03   & 0.05$\pm$0.03  & 1.88$\pm$0.02    & 2.50            & 2.70$\pm$0.02 & 2.4 & 0.56\\
CD$-$38 2151  & 0.56$\pm$0.02   &$-$2.42$\pm$0.09 & 2.87$\pm$0.04   & 0.90$\pm$0.08   & 1.13$\pm$0.02 & 0.90& 6.31 \\
HD 30443     & 2.16$\pm$0.05   & 0.76$\pm$0.05  & 2.20$\pm$0.03    &  --             & --            & 2.05 & -- \\
HD 50264      & 15.31$\pm$0.25  & 4.84$\pm$0.04  & $-$0.04$\pm$0.01 & 1.10            & 4.57$\pm$0.02 & 4.6 &0.16\\
HD 87080      & 4.58$\pm$0.07   & 2.54$\pm$0.03  & 0.88$\pm$0.1     & 1.30$\pm$0.02   & 3.62$\pm$0.01 & 3.7 &3.16\\
HD 87853     & 5.05$\pm$0.08   & 3.10$\pm$0.04  & 0.66$\pm$0.01    & 1.05            & 3.95$\pm$0.01 & 2.50 & 4.7 \\
HD 123701     & 2.94$\pm$0.06   & 0.69$\pm$0.04  & 1.62$\pm$0.02    & 2.50$\pm$0.02   & 3.08$\pm$0.01 & 3.5 & 0.59\\
HD 176021     & 19.59$\pm$0.5   & 3.94$\pm$0.01  & 0.32$\pm$0.002   & 0.85            & 4.08$\pm$0.005& 3.9 &0.1\\
HD 188985     & 12.79$\pm$0.05  & 3.84$\pm$0.01  & 0.36$\pm$0.004   & 1.20            & 4.38$\pm$0.005 & 4.3 & 3.98\\
HD 202020     & 8.30$\pm$0.07   & 3.89$\pm$0.02  & 0.34$\pm$0.01    & 1.10            & 4.19$\pm$0.005 & 4.4& 5.0\\
\hline
\end{tabular}
\end{table*}
}

\begin{figure}
\centering
\includegraphics[width=\columnwidth]{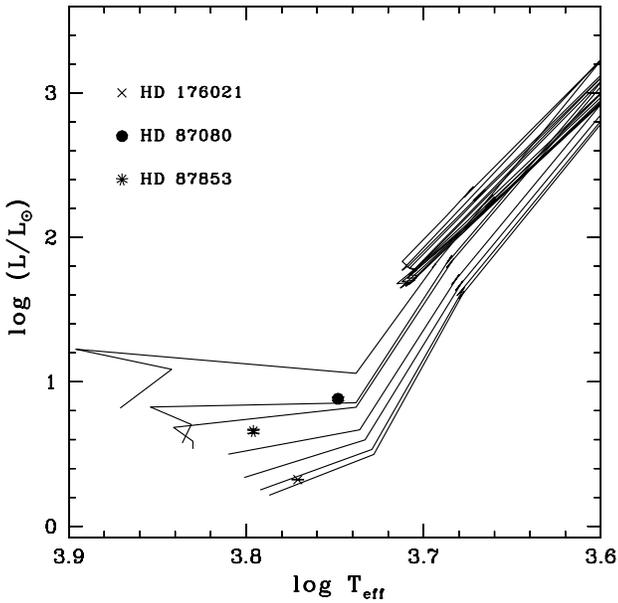}
\caption{The isochrone tracks for log(age) 10.05, 10.0, 9.9, 9.8, 9.55, 9.5, and  9.3 from bottom to top. }
\end{figure}

\begin{figure}
\centering
\includegraphics[width=\columnwidth]{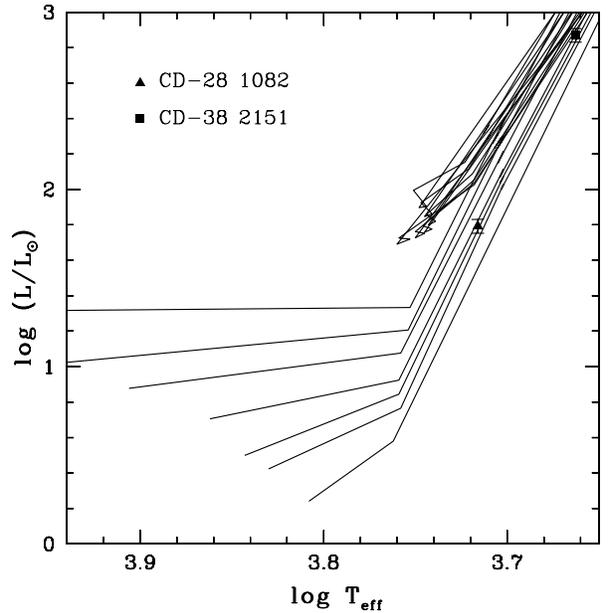}
\caption{The isochrone tracks for CD$-28$ 1082 and CD$-38$ 2151 corresponding to log(age) 
10.2, 10, 9.9, 9.8, 9.6, 9.4 and 9.25 from bottom to top. }
\end{figure}

\begin{figure}
\centering
\includegraphics[width=\columnwidth]{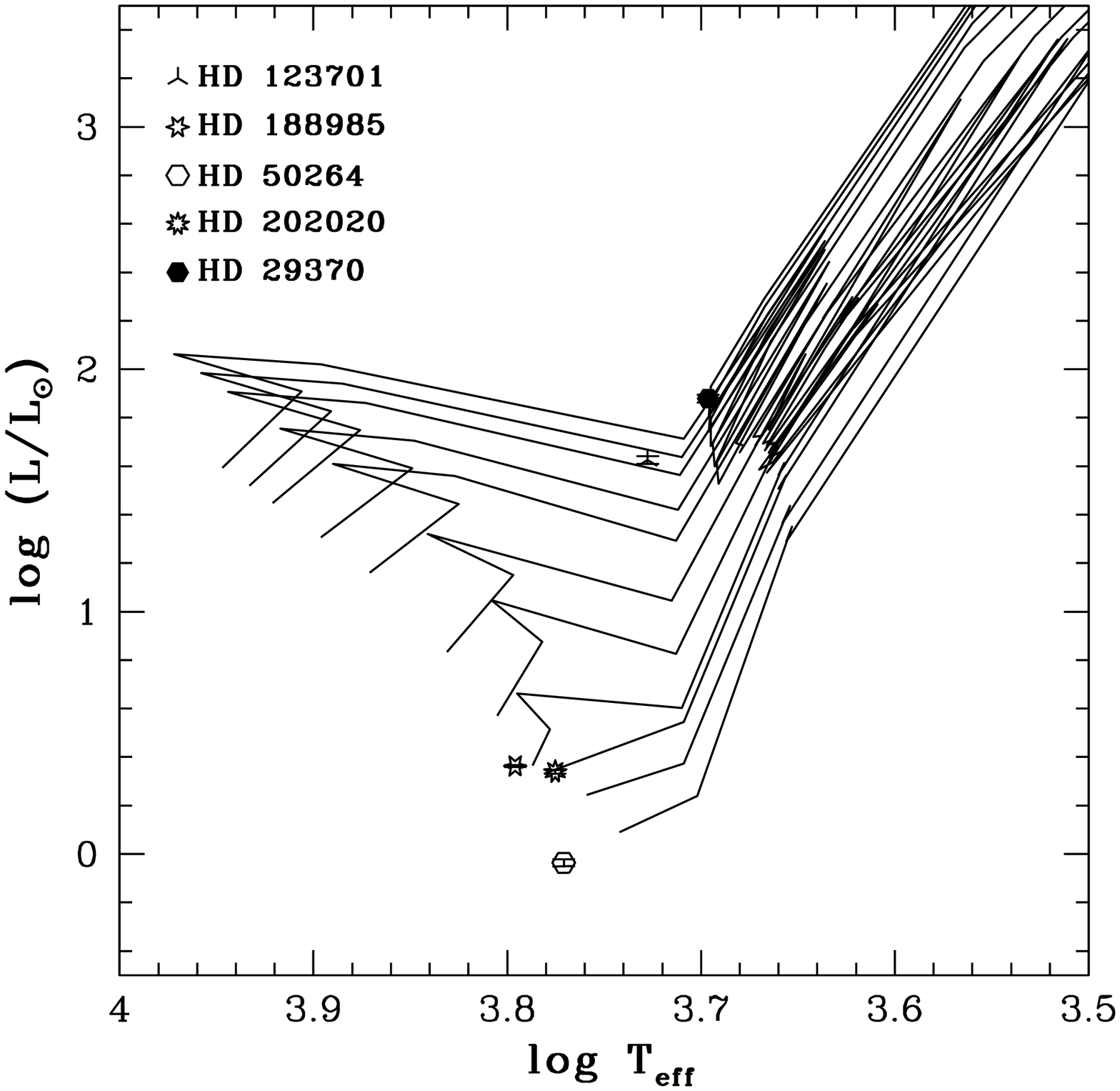}
\caption{The isochrone tracks for log(age) 10.20, 10.0, 9.70, 
9.60, 9.40, 9.20, 9.00, 8.90, 8.80, 8.75, and 8.70 are 
shown from bottom to top. }
\end{figure}

\section{Comparison with literature}
To verify our estimates measured from the  VBT spectra  we have 
analyzed anew the FEROS spectra of HD~87080 and  HD~188985  which 
were previously studied  by Allen \& Barbuy (2006a,b), and HD~123701 
and HD~29370 previously studied  by de Castro et al. (2016) and 
compared  their results with those obtained by us from VBT spectra. Our 
estimates of atmospheric  parameters and metallicity for these
stars  match closely with those obtained by Allen \& Barbuy (2006a,b) and 
de Castro et al. (2016) within the  error limits. As a check to our estimates 
we have randomly selected a few  lines, and performed spectrum 
synthesis calculation using   our estimates of atmospheric 
parameters as well as  the parameters obtained by Allen \& Barbuy (2006a)
 and de Castro et al. (2016) for the respective stars. The fits match 
quite well. 
As an  example we have shown the best fits obtained in a few cases  
(Fe I  5434.523 {\rm \AA}, Fe II  4620.521
{\rm \AA} and Ba II  5853.67 {\rm \AA} )  in  Fig. 8. 
We have adopted  our estimates for further  analysis of 
the objects. Table 5 presents a comparison of our estimates with those of
Allen \& Barbuy (2006), de Castro et al. (2016) and other literature
values. In Fig. 9, we have shown a comparison of VBT 
Echelle spectrum of CD$-$38 2151 with the FEROS spectrum. The spectral 
features match well. Fig. 10 illustrates a comparison of equivalent 
widths measured in FEROS spectrum with those measured on VBT spectrum;  
the measurements match closely. Such comparisons were performed for all 
the stars for which both VBT and FEROS spectra are available; here for 
illustration purpose we have shown the comparison  for  one case. 

\begin{figure}
\centering
\includegraphics[angle=0,height=7cm,width=8cm]{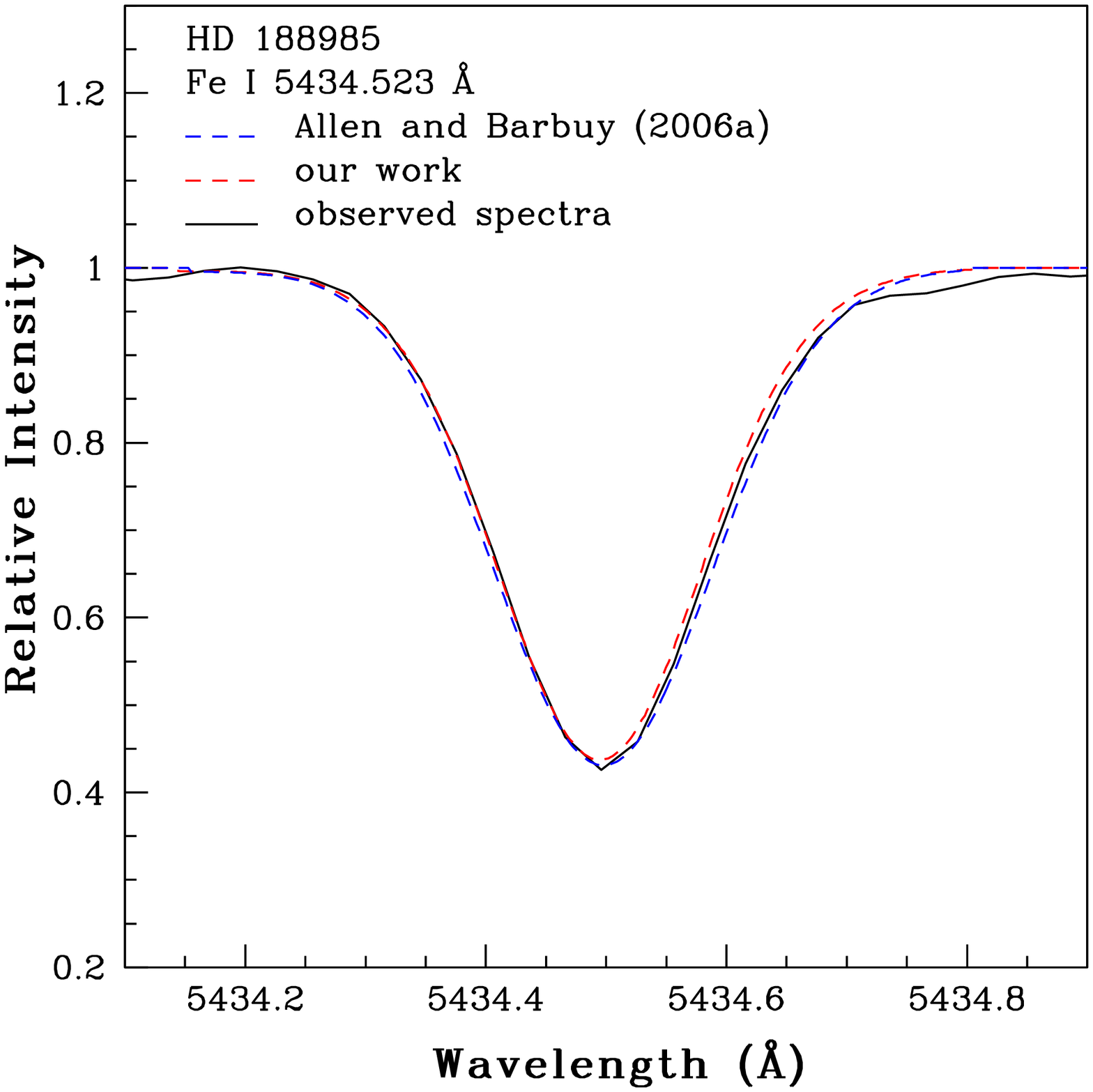}
\includegraphics[angle=0,height=7cm,width=8cm]{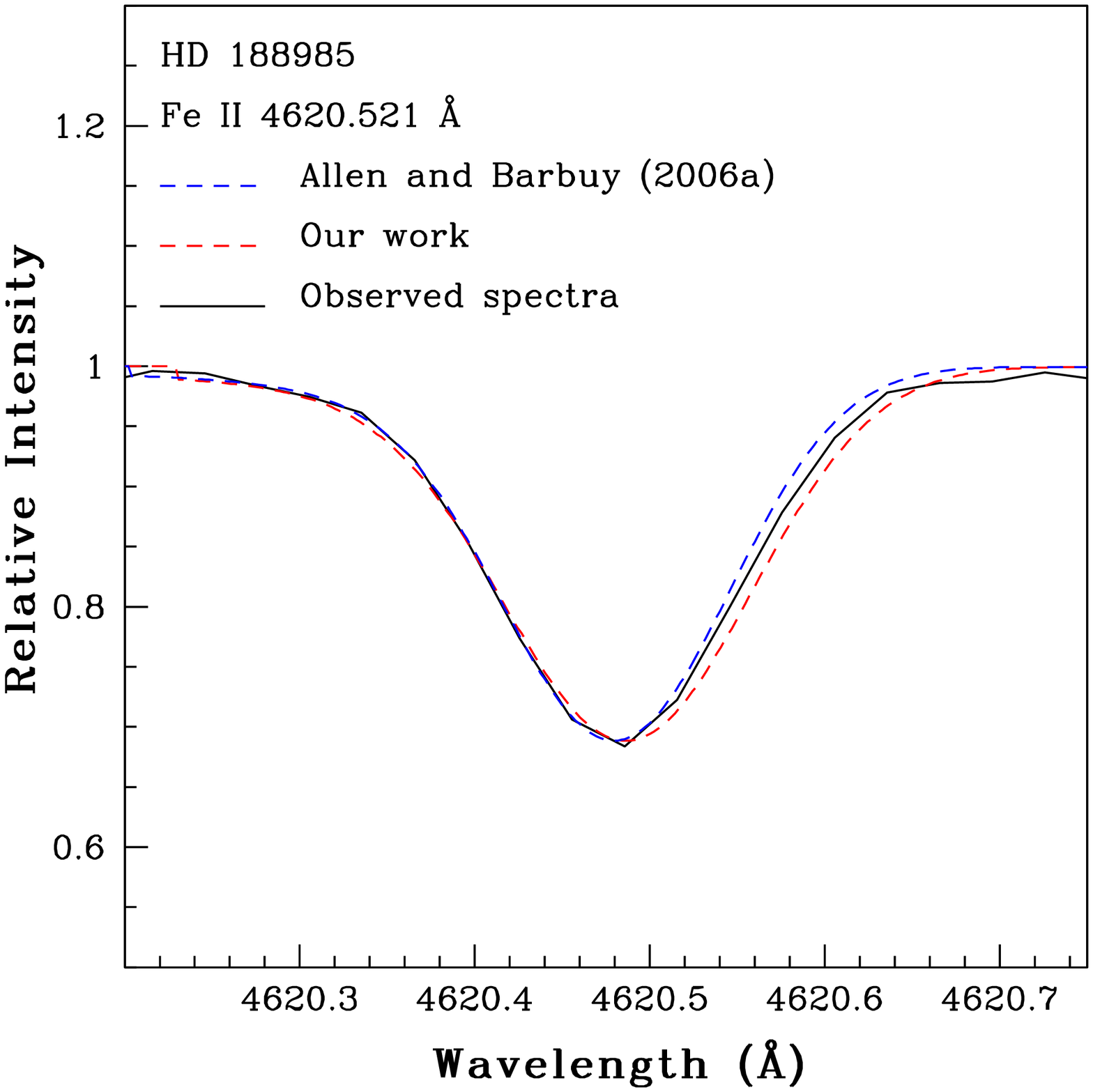}
\includegraphics[angle=0,height=7cm,width=8cm]{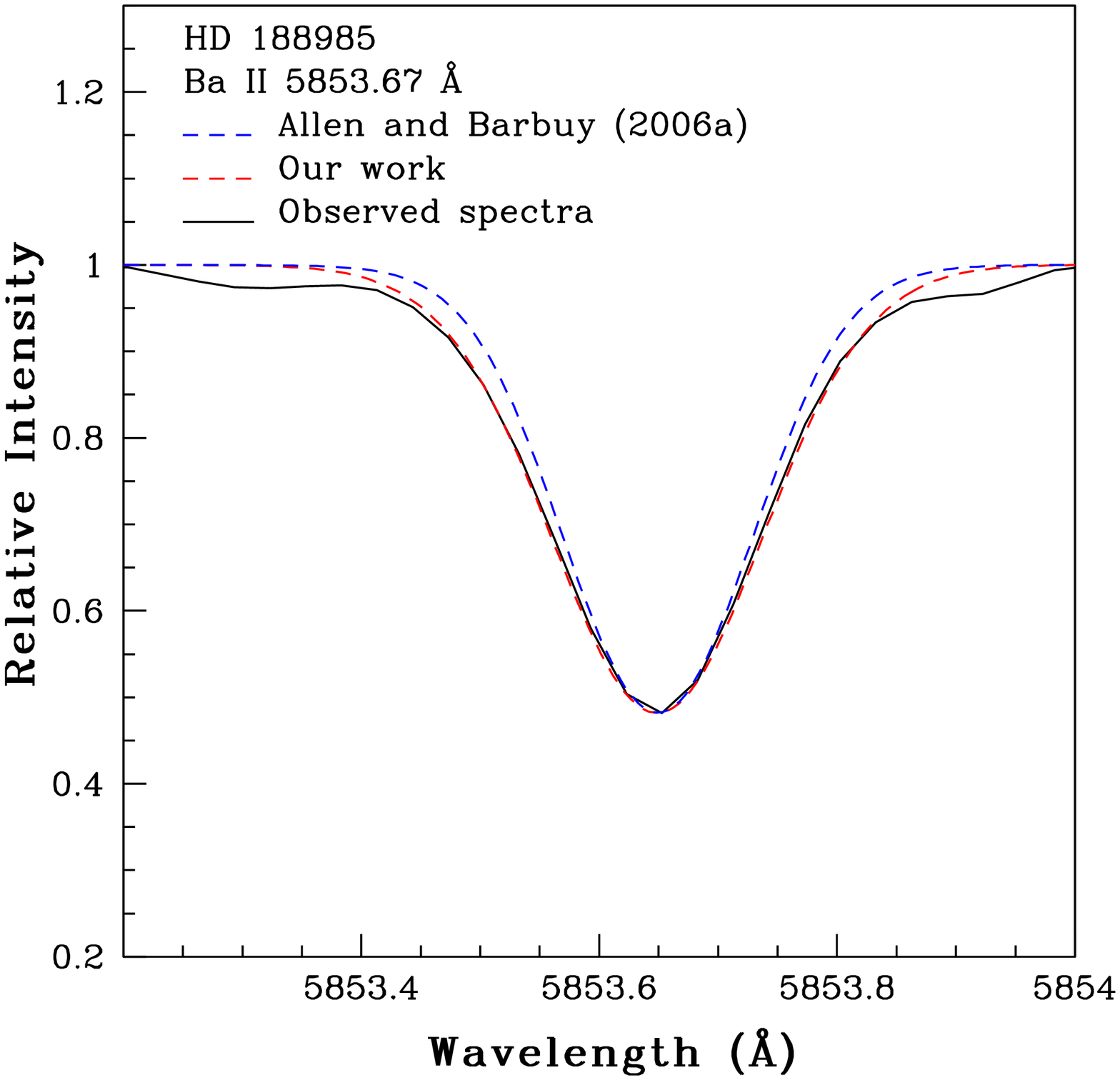}
\caption{  Synthesis of   Fe I  5434.523 {\bf  {\rm \AA}} (Top panel), 
Fe II 4620.521 {\bf  {\rm \AA}} (Middle panel), and  Ba II 5853.67 {\bf  {\rm \AA}} (Bottom panel). 
The best fits obtained  with our atmospheric parameters and those of 
Allen \&  Barbuy (2006a) are shown.
}
\end{figure}

\begin{figure}
\centering
\includegraphics[angle=0,height=7cm,width=8cm]{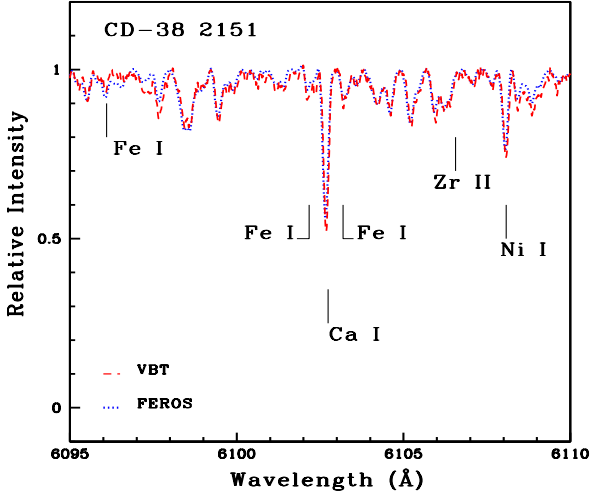}
\caption{ A comparison of the spectra of CD$-$38 2151 obtained with VBT 
with the  FEROS spectrum shows a good match.}
\end{figure}

\begin{figure}
\centering
\includegraphics[angle=0,height=7cm,width=8cm]{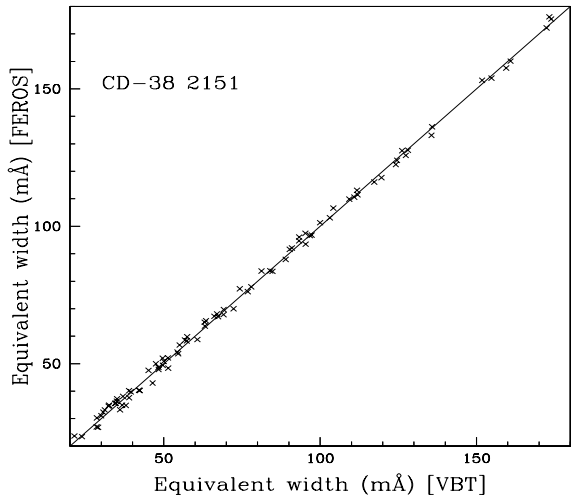}
\caption{ A comparison of the equivalent widths measured on the VBT 
spectrum of  CD$-$38 2151 compared with those measured on the FEROS spectrum.}
\end{figure}

{\footnotesize
\begin{table*}
{\bf Table 5: Comparison of our estimated stellar parameters with literature values }\\ 
\begin{tabular}{lccccccc}
\hline
Star        & T$_{eff}$ & log g &$\zeta$       & [Fe I/H]  &[Fe II/H] & Ref.  \\
            &    (K)    &  (cgs)     &(km s$^{-1}$) &           &           \\
\hline
HE~0308$-$1612& 4600    & 1.70  & 1.42         & $-$0.72   & $-$0.73  & 1 \\
CD$-$28 1082& 5200      & 1.90  & 1.42         & $-2.46$   & $-2.44$ & 1   \\
HD 29370    & 4970      & 2.40  & 1.92         & $-0.15$   & $-0.13$ & 1 \\
            & 4800      & 2.10  & 1.60         & $-0.25$   & $-0.28$ & 2 \\
HD 30443    & 4040      & 2.05  &2.70          & $-1.68$   & $-1.69$ & 1 \\
CD$-$38 2151& 4600      & 0.90  & 2.30         & $-2.03$   & $-2.03$ & 1 \\
            & 4700      & 1.5   & 3.00         & $-1.50^a$ &         & 3  \\
HD 50264    & 5900      &4.60   &0.62          & $-0.14$   & $-0.11$ & 1 \\
            & 5800      &4.20   & 1.00         & $-0.34$   & $-0.37$ & 4 \\
HD 87080    & 5600      & 3.74  & 1.10         & $-0.47$   & $-0.49$ & 1  \\  
            & 5600      & 4.0   & 1.2          & $-0.51$   & $-0.49$ & 4  \\
            & 5460      & 3.70  & 1.00         & $-0.49$   & $-0.44$ & 5 \\
HD 87853    & 6250      &2.50   & 1.63         & $-$0.72   &$-$0.73  & 1 \\
HD 123701   & 5340      & 3.5   & 1.44         & 0.05      & 0.07    & 1 \\
            & 5000      & 2.5   & 1.60         & $-0.44$   & $-0.46$ & 2 \\
HD 176021   & 5900      & 3.95  & 1.02         & $-0.62$   & $-0.65$ & 1 \\
            & 6000      & 4.00  & 2.00         & $-0.32$   & $-0.50$ & 6 \\
HD 188985   & 6250      & 4.30  & 0.60         & 0.01      & 0.01    & 1 \\
            & 6090      & 4.30  & 1.10         & $-0.25$   & $-0.30$ & 5\\
            & 5960      & 3.78  & 1.40         & $-$0.51   & $-$0.32 & 7 \\
HD 202020   & 5960      & 4.38  & 0.86         & 0.04      & 0.08    & 1 \\
            & 5600      & 4.00  & 1.20         & $-0.14$   & $-0.07$ & 8\\
\hline
\end{tabular}

\textit{Note.} References: 1. Our work, 2. de Castro et al. (2016), 3.  Vanture (1992), 
4. Pereira \& Junqueira (2003), 5. Allen \& Barbuy (2006a), 
6. Sneden \& Bond (1976), 7. North et al. (1994a,b) 8. Luck \& Bond (1991)\\
$^a$ Vanture (1992) value refers to  [Fe/H].

\end{table*}
}

\section{ABUNDANCE ANALYSIS}
Elemental abundances are measured using equivalent widths 
of good lines as well as using spectral synthesis calculation whenever 
applicable.  Lines due to different elements are identified by 
overplotting the Arcturus spectra on the spectra of our programme 
stars. A master line list  with the measured equivalent widths,
and line information such as lower excitation potential and log\,{gf} 
values of the lines  due to  different elements  considered for abundance
estimates  is prepared using  the Kurucz database.  Only symmetric 
and clean lines are considered for the measurement of equivalent 
widths. 
Abundances of 
light elements, C, N O, odd-Z element Na,  ${\alpha}-$ and  Fe-peak 
elements  Mg, Si, Ca, Sc, Ti, V, Cr, Mn, Co, Ni and Zn are measured. 
Among the neutron-capture elements we have estimated the  abundances 
of elements  Sr, Y, Zr, Ba, La, Ce, Pr, Nd, Sm, and Eu. 
For elements Sc, V, Mn, Ba, La and Eu, we have also used spectrum synthesis 
calculation taking into account their hyperfine structures. 

\par The abundance results are presented in Tables A1 through A4 and the 
 lines  used for the  abundance determination  are presented  in  Tables 
A5-A8.

\subsection{Carbon, Nitrogen, Oxygen}
Oxygen abundance is determined using spectrum synthesis calculation 
of the oxygen forbidden lines [OI] 6300.3 \r{A} and 6363.8 \r{A} lines 
for the objects 
CD$-$38 2151, HD~30443,  HD~123701, HD~202020. Some examples of spectrum 
synthesis fits are shown in  Fig. 11. In case of the objects 
HD~188985, HD~50264,  HD~87080 and HD~87853 where these lines could not be 
measured we have determined the oxygen abundance using spectrum 
synthesis calculation of  the triplet lines  around 7774 \r{A} 
region (Fig. 12).  
These lines are influenced by
 non-LTE effects. Non-LTE corrections are estimated using the equation
given  in  Bensby, Feltzing \& Lundstrom (2004). The NLTE effects on oxygen triplet
decreases in objects, such as  K giants  with higher gravities,
and it vanishes for
lines with equivalent widths below 45 {\rm m\AA}
(Eriksson \& Toft 1979).  As in most of the stars the O I triplet lines
are found to be stronger with equivalent width slightly larger  than
45 {\rm m\AA}, and hence  we have applied
non-LTE corrections to the estimated   abundances. Oxygen 
is found to be near solar in HD 188985, whereas it is underabundant in 
HD 202020 with [O/Fe] $\sim$ $-$0.33. For  HD 50264 and HD 87080 the 
estimated   [O/Fe]  ${\sim}$ 0.37.  Oxygen is moderately enhanced in 
HD~87853, HD 123701 and CD$-38$ 2151. In the rest of the objects  
the oxygen lines
are found to be blended and not usable for  abundance estimates.

\par Abundance of carbon is determined  from spectrum synthesis 
calculation of the wavelength region 5162 to 5176 \AA,  including the 
C$_{2}$ bandhead at 5165 \r{A}. In this wavelength region  there are 
no molecular  features due to
MgH, CN bands. Atomic  analysis of warm   K and M dwarfs,
shows that the  molecular absorption is not strongly pronounced in 
spectra of stars at  temperatures  T$_{eff}$ $>$ 3500 K (Woolf  and 
Wallerstein 2005).
 The molecular lines used for the synthesis 
is taken from Kurucz database. Spectrum synthesis fits for a few 
programme stars are shown in Fig. 13.

\par CD$-28$ 1082, HD~30443 and CD$-38$ 2151 show 
carbon enhancement with [C/Fe] $>$ 1 while HE~0308$-$1612 and  HD~176021 
are moderately enhanced with [C/Fe] $\sim$ 0.78 and 0.52, respectively. 
Abundance of carbon could not be estimated for HD~87853.  In the 
remaining stars,  [C/Fe] ${\le}$ 0.23  or near solar. 
\par The carbon isotopic ratio $^{12}$C/$^{13}$C is determined from the 
spectrum synthesis calculation of the
 CN band at 8005 \r{A} (Fig. 14).  We could estimate $^{12}$C/$^{13}$C 
ratios for six objects HE~0308$-$1612, CD$-28$ 1082, HD 29370, HD~30443,  
 CD$-38$ 2151 and HD~123701;
the values are 15.6, 16, 7.4, 9.3, 11.2 and 10, respectively. In the 
rest of the stars 
this band is  found to be either too weak or absent. 
While in CD$-38$ 2151, HD~30443 
and HD 202020,   C/O is found to be  $>$ 1 as normally seen in  CH stars, 
HD~50264, HD~87080, HD~123701 and  HD~188985 show C/O $<$ 1.

\begin{figure}
\centering
\includegraphics[width=\columnwidth]{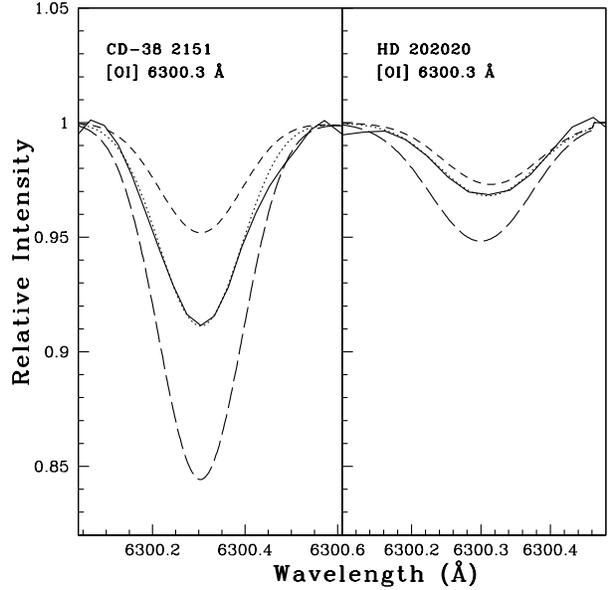}
\caption{ Synthesis of [OI] line around 6300 {\rm \AA}. Dotted line represents synthesized spectra and the solid line indicates the observed spectra. Short dashed line represents the synthetic spectra corresponding to $\Delta$[O/Fe] = -0.3 and long dashed line is corresponding to $\Delta$[O/Fe] = +0.3}
\end{figure}

\begin{figure}
\centering
\includegraphics[width=\columnwidth]{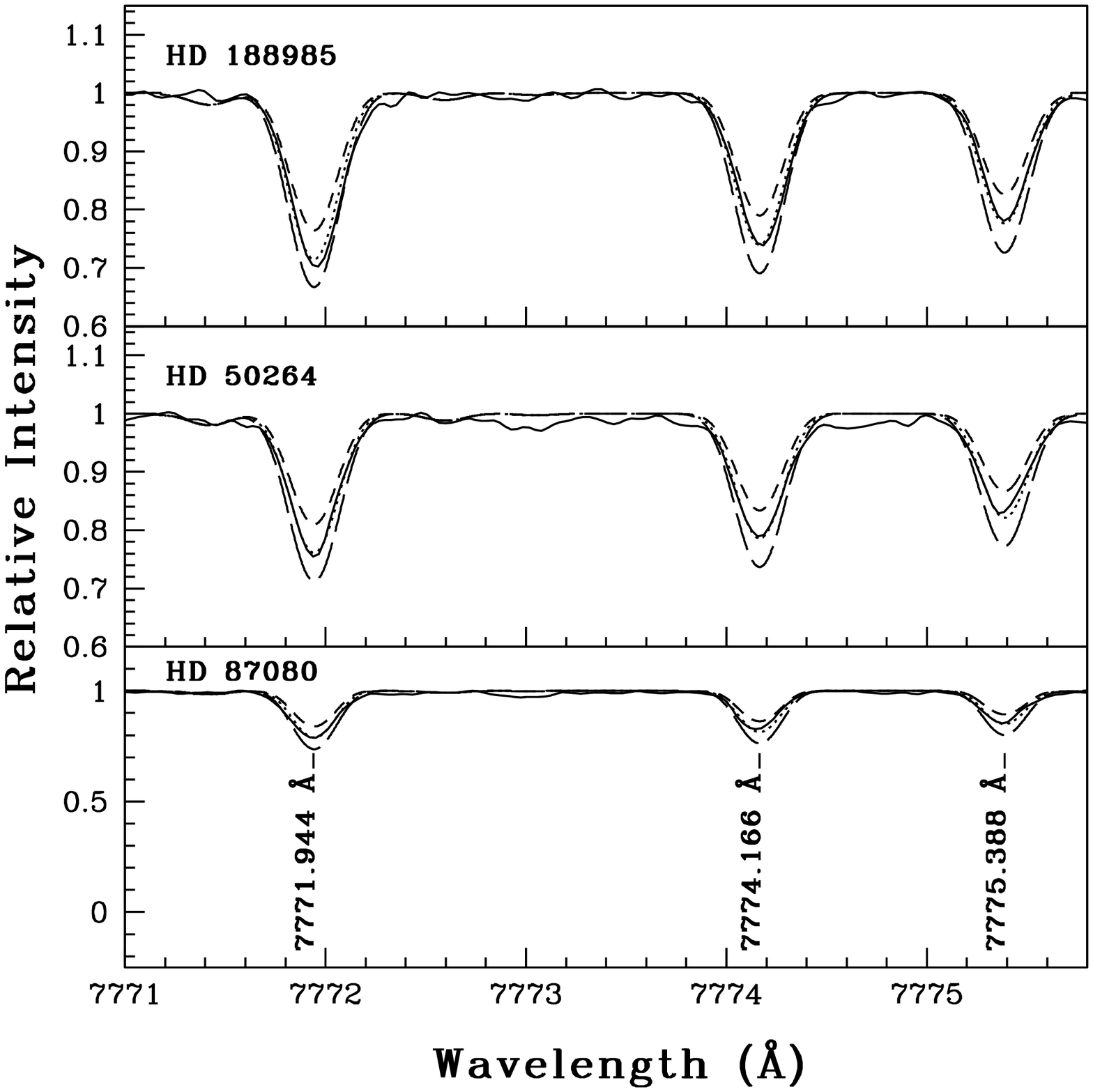}
\caption{ Synthesis of O I triplet. Dotted line represents synthesized spectra and the solid line indicates the observed spectra. Short dashed line represents the synthetic spectra corresponding to $\Delta$[O/Fe] = -0.3 and long dashed line is corresponding to $\Delta$[O/Fe] = +0.3}
\end{figure}

\begin{figure}
\centering
\includegraphics[width=\columnwidth]{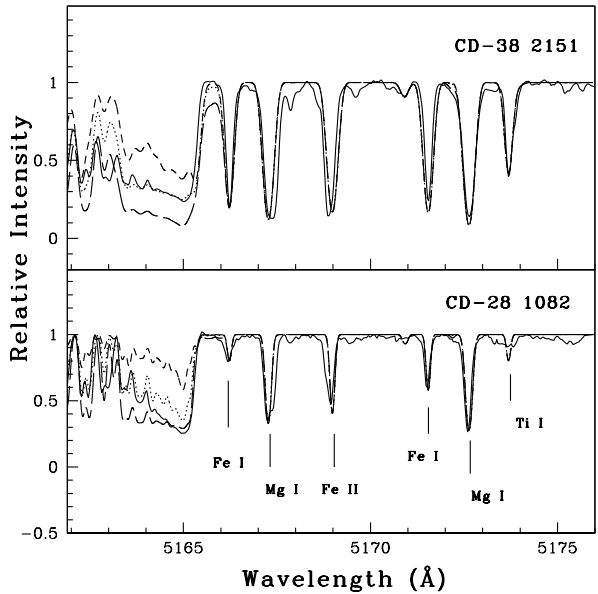}
\caption{ Synthesis of C$_{2}$ band around 5165 {\rm \AA}. Dotted line 
represents synthesized spectra and the solid line indicates the 
observed spectra. Short dashed line represents the synthetic spectra 
corresponding to $\Delta$ [C/Fe] = -0.3 and long dashed line is 
corresponding to $\Delta$[C/Fe] = +0.3}
\end{figure} 

\begin{figure}
\centering
\includegraphics[width=\columnwidth]{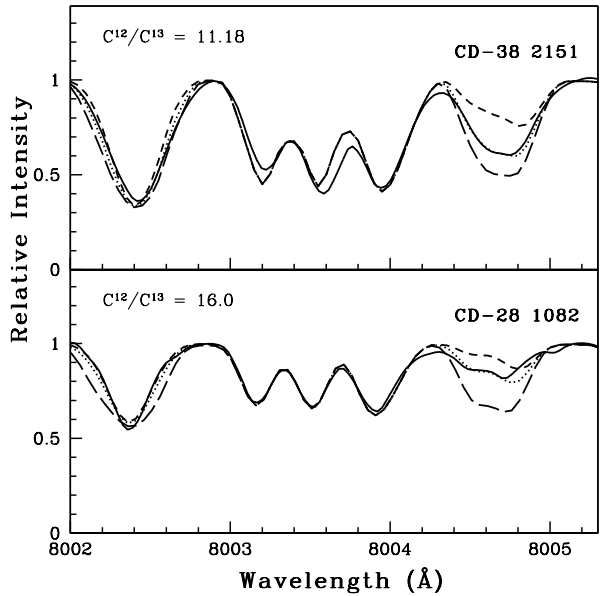}
\caption{ Synthesis of CN band around 8005 {\rm \AA}. Dotted line 
represents synthesized spectra and the solid line indicates the 
observed spectra and the synthetic spectra corresponding to 
$^{12}$C/$^{13}$C $\simeq$ 12 (short dashed) and 1 (long dashed)}
\end{figure}

\par 
Nitrogen is estimated using the spectrum  synthesis calculation of the 
CN band at 4215 \r{A} (Fig. 15). Nitrogen is enhanced in CD$-28$ 1082 
and CD$-38$ 2151 with [N/Fe] $\sim$ 2.73 and 1.40, respectively. The 
rest of the stars show moderate enhancement in  nitrogen relative to 
iron. HE~0308$-$1612 and  HD~29370 show  near solar values. Abundance 
of Nitrogen could not be estimated for HD~87853.

\begin{figure}
\centering
\includegraphics[width=\columnwidth]{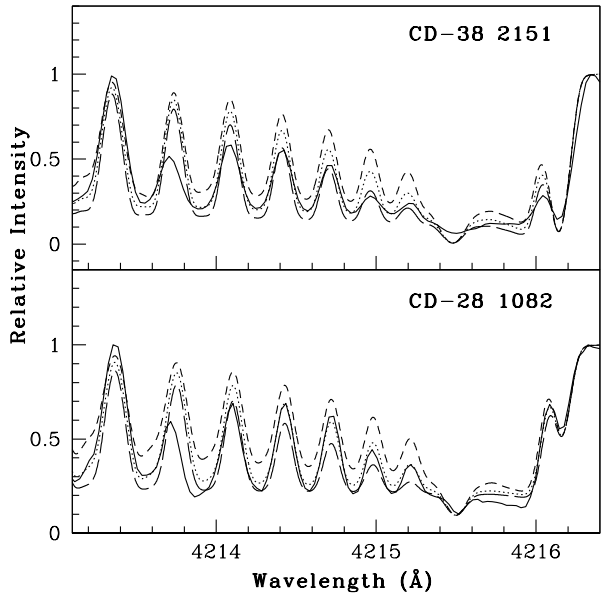}
\caption{ Synthesis of CN band around 4215 {\rm \AA}. Dotted line 
represents synthesized spectra and the solid line indicates the 
observed spectra. Short dashed line represents the synthetic spectra 
corresponding to $\Delta$[N/Fe] = -0.3 and long dashed line is 
corresponding to $\Delta$[N/Fe] = +0.3}
\end{figure}

\subsection{Na, Mg, Si, Ca, Sc, Ti, V}
The estimated  Na abundances lie in the range 0.07-0.78. 
While five objects, HD 29370, HD 87080, HD 123701, HD 176021 and 
HD 202020 show solar values. CD $-$28 1082 shows an enhancement 
with [Na/Fe] ${\sim}$ 0.78. Such an enhancement is not uncommon 
for stars at the  metallicity of this star (Aoki et al. 2007). The 
lines used for abundance estimates are presented in Tables A7 and A8.
 We could not estimate  the abundance of sodium  in CD$-38$ 2151, 
HD~50264 and HD~188985 as no suitable lines were found  for 
abundance estimates.

\par 
Abundance of Mg is determined  using lines Mg I 4571.10, 4730.03,
5172.684, 5528.405, and 5711.09 \AA\, whenever possible using spectrum
synthesis calculations.  For all the stars, the lines used for 
abundance analysis have equivalent widths in the range
20 - 180 m\AA\,. In the case of HE 0308-1612 since only two Mg lines, 
Mg I 5528.405, and 4702.99 \AA\, are found to be useful for abundance 
analysis, we have used both the lines although Mg I 5528.405 has
equivalent width slightly larger than 180 m\AA\, (i.e., 190.6 m\AA\,).
 In CD$-28$ 1082, Mg is found  to be enhanced 
with [Mg/Fe] ${\sim}$ 0.45. In CD$-38$ 2151, this value is slightly
higher with [Mg/Fe] $\sim $ 0.63. HD~87853 also shows a similar value 
with [Mg/Fe] $\sim $ 0.68. A value of 0.38 is obtained for HD 176021.
For the rest of the objects we have obtained near solar values 
(Tables A1 - A4).

\par Abundance of Si is determined using lines, Si I 6145.02, 6155.13 
and 6237.32 \AA\, whenever possible.
While HD~29370 and HD~87080  show a mild deficiency with
[Si/Fe] = $-$0.36 and $-$0.44 respectively.  CD$-38$ 2151 and HD~30443
give values of [Si/Fe] = 1.62 and 0.82 respectively. In the rest 
of the objects abundance of Si could not be determined as no good
lines were found for abundance determination. 

\par   Ca shows a mild enhancement  of [Ca/Fe] = 0.27, 0.37, 0.29
in  CD$-28$ 1082, HE~0308$-$1612,  HD~87853.
Ca is also moderately enhanced in HD~30443 and CD$-38$ 2151 with values
$\sim$ 0.51 and 0.58 respectively. HD~176021 shows a near solar value, and
for the remaining six objects the  [Ca/Fe] lies in the range from 
 $-$0.25 to $-$0.08.

\par  Abundance of Sc  determined using the equivalent width 
measurements of the lines Sc II 4431.35,  5239.81\AA\, and 5526.79 \AA\
give near solar values for CD$-$38 2151 and HD~29370 with [Sc/Fe] ${\sim}$
0.18 and 0.19 respectively.
Spectrum synthesis of Sc II 6245.63 \r{A}  is used to estimate  the 
abundance of Sc in HE~0308$-$1612, that gives   [Sc/Fe] ${\sim}$ 0.53,
and near solar values for  HD~30443, HD~50264, 
HD~87080, HD~123701, HD~176021, and HD~202020. 
The objects HD~188985 and HD~87853 show  mild deficiencies  with
[Sc/Fe] ${\sim}$ $-$0.21 and  $-$0.39 respectively. The 
hyperfine structure of Sc is taken from  Prochaska and McWilliam (2000). 
Abundance of Sc could not be measured in  
CD$-28$ 1082 as no good lines were found for abundance determination.

\par Abundance of Ti is measured using the equivalent width measurements
of several Ti I lines (Tables A7, A8). [Ti/Fe] covers a range 
$-$0.28 to 0.32, in these objects except for  CD$-38$ 2151 where 
[Ti I/Fe] is 0.71. The estimates of [Ti II/Fe] covers a range
$-$0.28 to +0.42.
We could not estimate abundance of Ti for  CD$-28$ 1082 as no good
lines were found.

\par 
 The abundance of V  measured from 
V I 6119.52, and 6251.83 \AA\, lines give a very high value with
[V/Fe] ${\sim}$ 1.16 for CD$-38$ 2151. The reason for such 
a high value is not understood. For HD~29370 and  HD~123701 
we get [V/Fe] 
values as 0.29 and 0.12.  The spectrum synthesis calculation 
of the lines V I 5727.048, 5727.652, and  5737.06 \r{A}  taking 
hyperfine splitting  into considerations returns [V/Fe] values 
0.29, $-$0.64,  $-$0.43 and $-$0.18 for HE~0308$-$1612, HD~30443,  
HD~176021 and HD~202020 respectively. Hyperfine structure is taken 
from Prochaska \&
McWilliam (2000). The abundance of V could not be estimated for the
remaining four objects. 
 
\subsection{Cr, Co, Mn, Ni, Zn}
The abundance of Cr is derived using equivalent width measurements of 
several Cr I lines; [Cr I/Fe] ranges from $-$0.39 to 0.30 in these objects.
We could also estimate abundance of Cr using Cr II lines in 
 HD~29370, HD~50264, HD~87080, HD~87853 and HD~176021; 
[Cr II/Fe] ranges from $-$0.18 to 
0.03 in these objects. Cromium abundance could not be estimated for  
CD$-$28 1082 and HD~30443. 

\par The abundance of Mn is determined using the spectrum  synthesis 
calculation  of  the Mn I  6013.51 \r{A} considering
 the hyperfine structure  from Prochaska \& McWilliam (2000). 
CD$-28$ 1082 shows  a large enhancement of Mn with [Mn/Fe] $\sim$ 1.48. 
HD~30443 shows near solar value.
HD~29370 and HD~176021 give a mild enhancement with [Mn/Fe] values  
0.32 and 0.17 respectively.  Spectrum synthesis calculation of 
Mn I 4041.355 \AA\, gives a value $-$0.37 for HD~87853. Abundance of Mn 
derived using equivalent width measurements of  lines, Mn I 4739.087, 
4766.418 and 4783.427 \AA\, is
found to be ${\sim}$ 0.29 with respect to iron for HE~0308$-$1612.
In the rest of the stars [Mn/Fe] ranges from
$-$0.38 to $-$0.2.

\par Abundance of Co is measured using the Co I  4121.31, 
4792.85, 4813.47, 5483.34 and 6632.43 \AA\, lines whenever possible.
For  CD$-38$ 2151, HD~30443, HD~87853  and HD~ 123701, abundance of Co 
is  found to be  near solar.  For  HD~176021 and HD~188985  [Co/Fe] 
values are found to be mildly enhanced with [Co/Fe] ${\sim}$  0.29 
and 0.33  respectively.  For all other stars [Co/Fe] ranges from $-$0.01 
to $-$0.11. The abundance of Co could not be estimated for 
HE~0308$-$1612 and CD$-$28 1082  as no good lines could be measured.

\par Abundance of Ni is measured using several Ni I lines (Tables A7 and A8). 
[Ni/Fe]  ranges from 0.59 to  $-$0.13. 
The abundance of Ni could not be estimated for CD$-$28 1082. 

\par Abundance of Zn measured using only one line at 4810.53 \AA\,  
gives near solar values for   HD~29370,  HD~87080, HD~87853 and 
HD~123701. For stars HD~30443, HD~50264, HD~87853,  HD~176021, 
HD~188985, and HD~202020 the [Zn/Fe] values are found to  lie in 
the range   0.21-0.40.  HE~0308$-$1612 gives a value $-$0.19.
 We could not estimate the abundance of Zn for 
CD$-$28 1082  and  CD$-$38 2151 as no good lines are available in their
spectra.

\subsection{Sr, Y, Zr}
Abundance of Sr is determined from  spectrum synthesis calculation 
of Sr I  4607.33 \r{A}. Strontium is enhanced in most of the 
stars with  [Sr/Fe] $>$ 1. In HD~29370 and HD~30443, Sr is moderately 
enhanced with [Sr/Fe] $\sim$ 0.82 and 0.74 respectively. A few examples 
of the spectrum synthesis fits for Sr is shown  in Fig. 16. Abundance 
of Sr could not be estimated for HD~87853.

\par Abundance of Y is measured using several lines of Y II (Tables A7, A8).
In HD~29370 and CD$-38$ 2151 yttrium is enhanced with [Y/Fe] values 0.87 
and 0.57 respectively. In HD~87853  yttrium  is underabundant with 
[Y/Fe] ${\sim}$ $-$0.48. In all other stars yttrium is enhanced with 
[Y/Fe] $>$ 1.0; with highest enhancement seen in CD$-28$ 1082 with 
[Y/Fe] ${\sim}$ 1.61.

\par  Abundance of Zr is determined using the measured equivalent width of 
Zr II 4210.631 \AA\,. Zr is found to be enhanced in  HD~29370, CD$-38$ 2151, 
HD~123701 and HD~202020  with [Zr/Fe] values  1.64, 2.0, 0.83 and 0.48 
respectively.
Spectrum synthesis calculation of Zr I  6134.59 \AA\, gives 
[Zr/Fe] ${\sim}$ 1.60 for HD~30443.
In the rest of the objects the abundance of Zr could not be estimated
as no good lines of Zr was found. 

\begin{figure}
\centering
\includegraphics[width=\columnwidth]{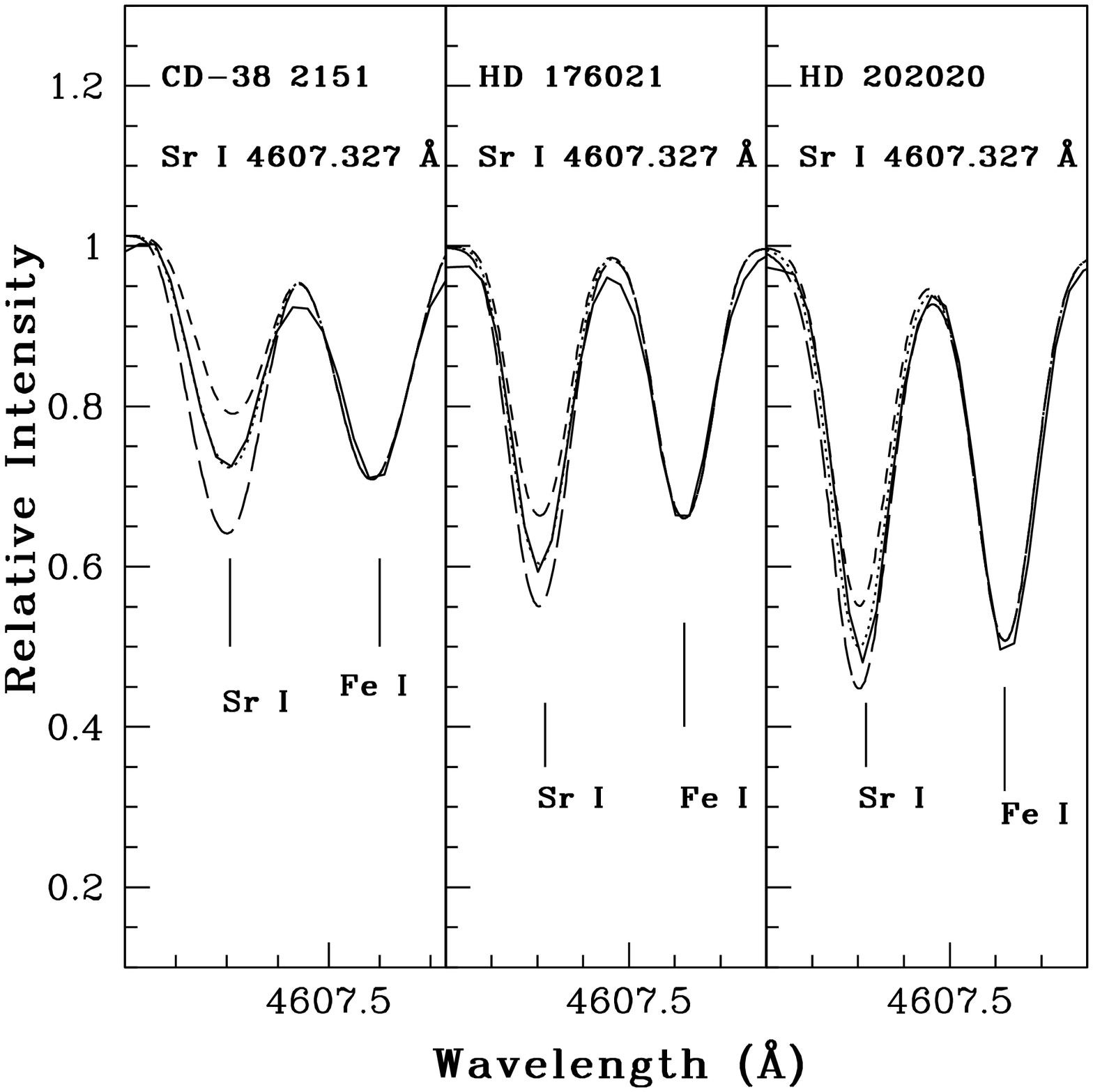}
\includegraphics[width=\columnwidth]{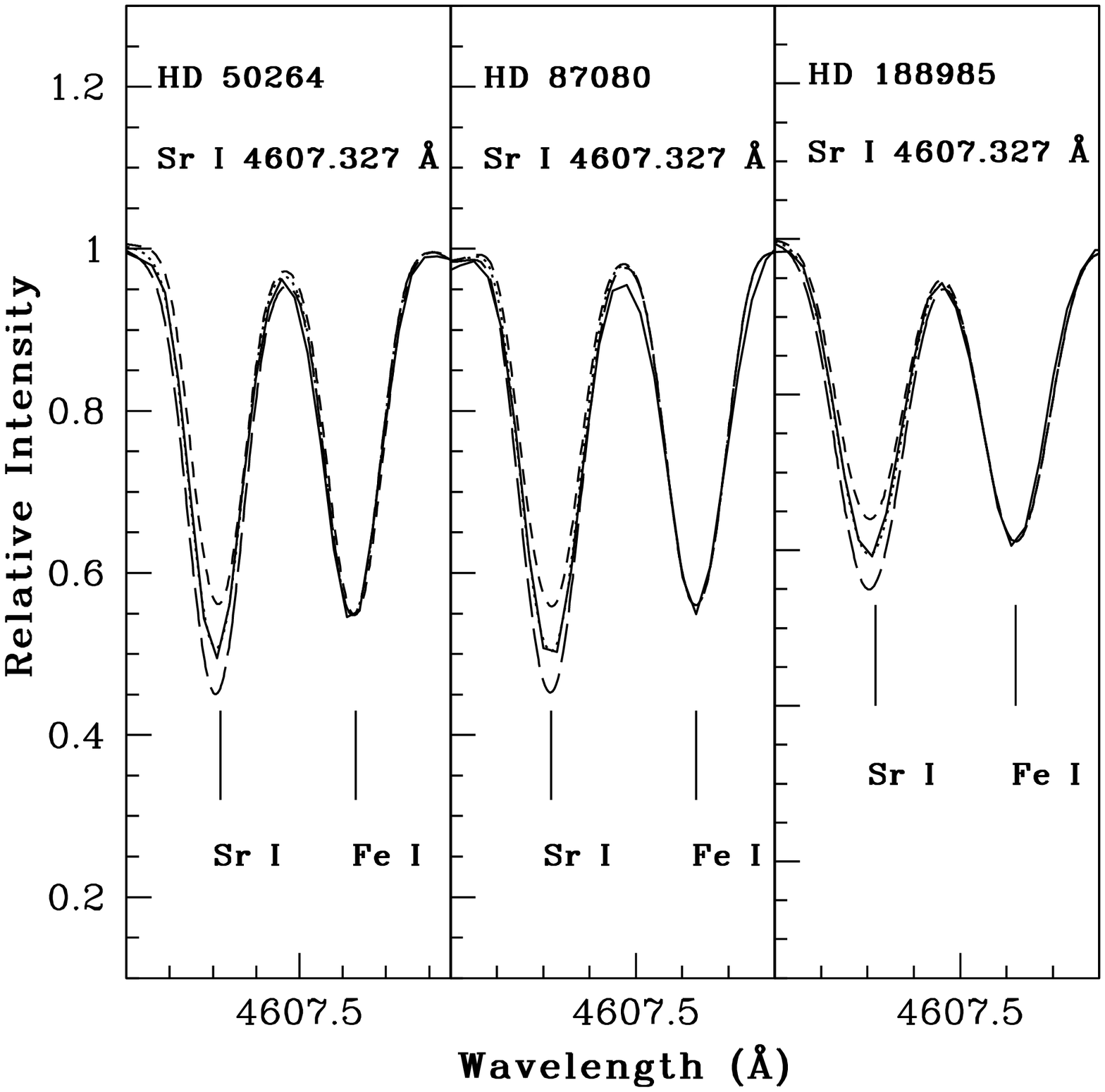}
\caption{ Synthesis of Sr I around 4607 {\rm \AA}. Dotted line represents synthesized spectra and the solid line indicates the observed spectra. Short dashed line represents the synthetic spectra corresponding to $\Delta$[Sr I/Fe] = -0.3 and long dashed line is corresponding to $\Delta$[Sr I/Fe] = +0.3}
\end{figure}

\begin{figure}
\centering
\includegraphics[width=\columnwidth]{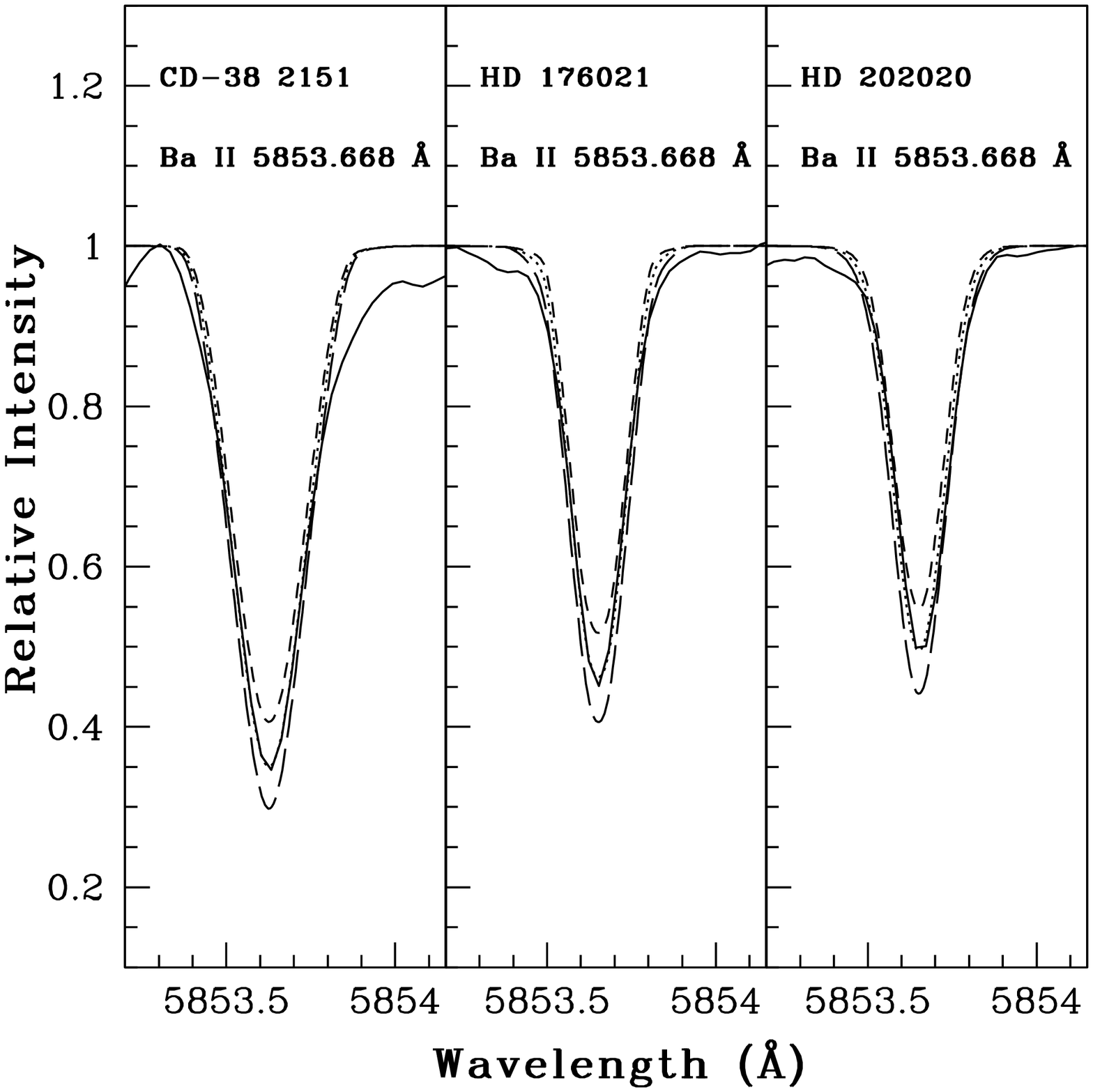}
\includegraphics[width=\columnwidth]{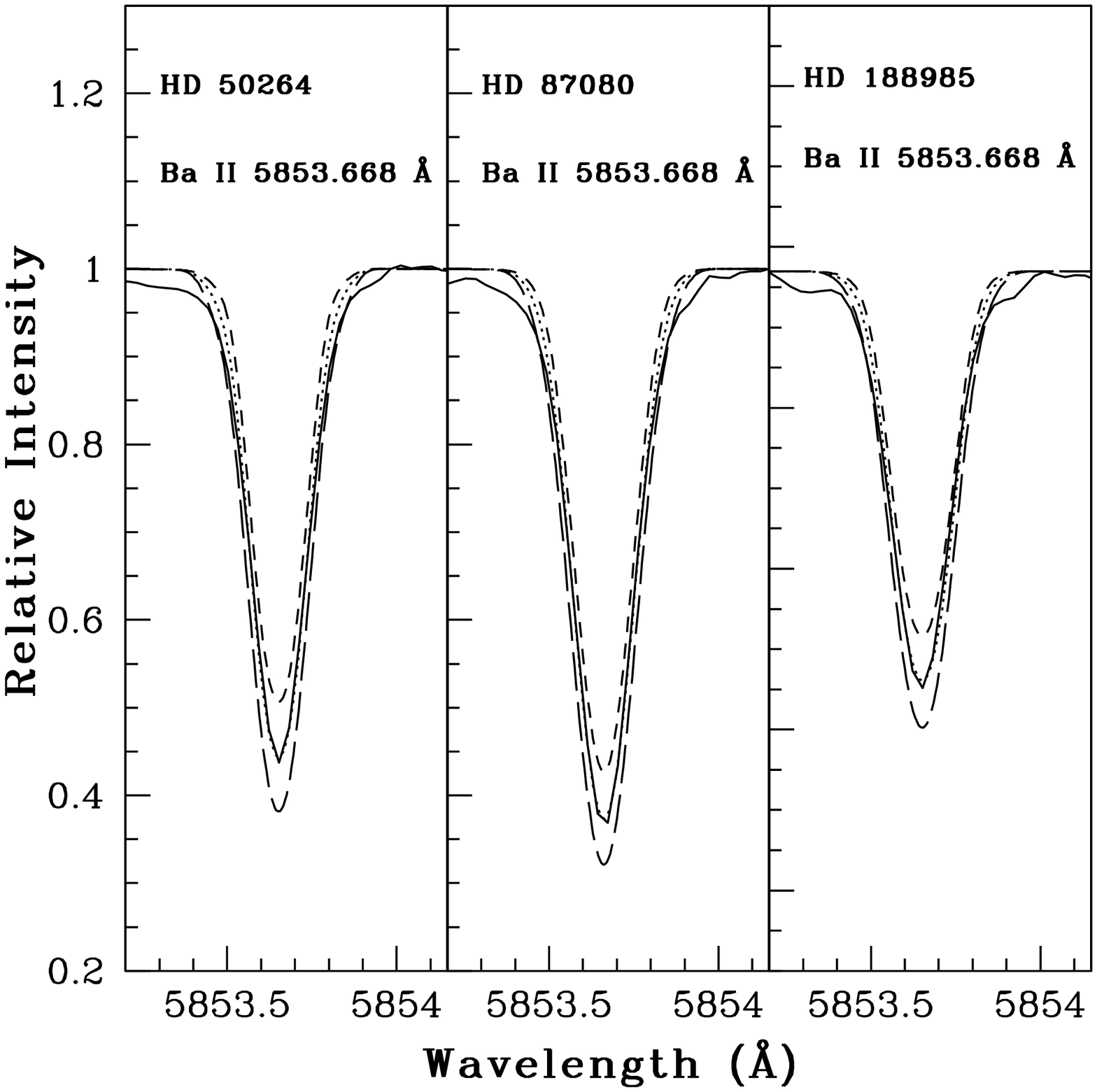}
\caption{ Synthesis of Ba II around 5835 {\rm \AA}. Dotted line represents synthesized spectra and the solid line indicates the observed spectra. Short dashed line represents the synthetic spectra corresponding to $\Delta$[Ba II/Fe] = -0.3 and long dashed line is corresponding to $\Delta$[Ba II/Fe] = +0.3}
\end{figure}

\subsection{Ba, La, Ce, Pr, Nd, Sm, Eu}
The abundance of
barium is determined from the spectrum synthesis calculation of the Ba II 
 5853.668 \r{A} (Fig. 17) considering the hyperfine structure  from 
McWilliam (1998).  For HD~30443 we have used spectrum synthesis calculation 
of Ba II  6141.713 \AA\,. Except for HD~87853 where barium is 
underabundant, all  of our programme stars show Ba 
enhancement with [Ba/Fe] in the  range   0.8-2.1. 

 \par Abundance 
of La is also determined from spectral synthesis calculation of La II
 4921.778 \AA\, with the 
hyperfine structure taken from Jonsell et al. (2006). La is 
overabundant in most of the stars with [La/Fe] ranging from 0.75 to 1.75. 
Abundance of La could not be
estimated for HD~30443 and HD~87853.

\par The abundance of cerium is estimated using the equivalent width 
measurements of several Ce II lines (Tables  A7, A8). The [Ce/Fe] values
are found to lie in the 0.8-1.97. 
We could not estimate Ce abundance for HD~87853.

\par The abundance of Pr is measured using the equivalent width 
measurements of  Pr II lines (A7, A8) that show enhancement in 
HE~0308$-$1612, CD$-$28 1082, HD~30443,  CD$-$38 2151, HD~123701, and
HD~87080 with [Pr/Fe] in the range 1.52-2.67.
Abundance of Pr could not be measured for  HD~29370, HD~50264, HD~87853, 
HD~176021, HD~188985 and HD~202020.

\par Abundance of Nd is measured using equivalent width measurements of
several Nd lines (Tables A7, A8).  Neodimium is found to be 
enhanced with [Nd/Fe] $>$ 1,  in all the stars except for HD~87853 and 
HD~202020.

\par  The abundance of Sm is measured using equivalent width 
measurements of a maximum of five  samarium lines. Sm is enhanced 
in CD$-28$ 1082 and HD~30443  with [Sm/Fe] ${\sim}$ 2.29 and 2.24
respectively. HE~0308$-$1612, HD~29370, 
CD$-38$ 2151, HD~87080, HD~123701 and HD~176021  show 
enhancement with [Sm/Fe] in the range 1.41-1.68.
Abundance of Sm could not be estimated for HD~50264, HD~87853, HD~188985, 
and HD~202020 as no good lines of Sm were available for abundance 
estimation.

\par Eu abundance is determined for six objects using spectrum
synthesis calculation of  lines Eu II 4129.71 \r{A}, 6437.63 \r{A}, 
and 6645.11 \r{A} considering hyperfine  structure from Worley 
et al. (2013). CD$-28$ 1082 shows  enhancement with 
[Eu/Fe] ${\sim}$ 2.07 measured using only one line Eu II 6437.63 \r{A}. 
In HD~123701, this line shows a marginal  enhancement with 
[Eu/Fe] = 0.15. For HE~0308$-$1612 and HD~87080 we have used the 
Eu II 6645.11 \r{A}
that gives [Eu/Fe] ${\sim}$ 0.36. Abundance of Eu estimated
using the line Eu II 4129.71 \AA\, gives [Eu/Fe] values 0.38, 0.10 and 
0.07 for HD~176021, 188985 and HD~202020.
Abundance of Eu measured using equivalent width 
measurement of Eu II  4205.042 \AA\, gives 
[Eu/Fe] ${\sim}$ 0.34 for HD~87853. 
Abundance of Eu could not be
estimated for   HD~29370, HD~30443, CD$-38$ 2151, and HD~50264
as no good lines are available for abundance estimation.

We have estimated [ls/Fe], [hs/Fe] and [hs/ls], where ls stands for
the light s-process elements (Sr, Y, Zr) and hs stands for the heavy 
s-process (Ba, La, Ce, Nd, Sm) elements; these values are presented 
in Table 6.

{\footnotesize
\begin{table*}
{\bf Table 6: Estimates of [Fe/H], [ls/Fe], [hs/Fe], [hs/ls] and C/O}\\ 
\begin{tabular}{lccccc}
\hline                       
Star name    & [Fe/H]  & [ls/Fe] & [hs/Fe] & [hs/ls] & C/O\\ 
\hline
HE~0308$-$1612&$-$0.73 & 1.11    & 1.59    &  0.48  & --  \\    
CD$-28$ 1082 & $-2.45$ & 1.74    & 2.39 & 0.65 & -- \\
HD 29370     & $-0.14$ & 1.14    & 1.28 & 0.14 & --  \\  
HD 30443     & $-1.68$ & 1.24    & 2.01 & 0.77  &  1.02 \\
CD$-38$2151  &$-2.03$  & 1.24    & 1.28 & 0.04 & 2.95  \\
HD 50264     &$-0.13$  & 1.54    & 1.43 &$-0.11$ & 0.41  \\
HD 87080     &$-0.48$  & 1.48    & 1.64 & 0.16 & 0.29 \\
HD 87853     &$-$0.73  & $-$0.48 & 0.26 & 0.74 & --  \\
HD 123701    & 0.06    & 1.19    & 1.62 & 0.43 & 0.15 \\
HD 176021    &$-0.64$  & 1.16    & 1.38 & 0.22 & -- \\
HD 188985    &0.01     & 1.60    & 1.40 & $-0.20$& 0.89 \\
HD 202020    & 0.06    & 1.10    & 0.77 & $-0.33$& 1.99 \\            
\hline
\end{tabular}
\end{table*}
}

\section{KINEMATIC ANALYSIS}
Space velocity of the programme stars are calculated using  the method 
described  in Bensby, Feltzing \& Lundstrom (2003).  The space velocity with respect 
to the Local Standard of Rest (LSR) is  given by  \\

 \begin{center}
  $(U, V, W)_{LSR} =(U,V,W)+(U, V, W)_{\odot}$ km/s.
  \end{center}

 where, $(U, V, W)_{\odot} =(11.1, 12.2, 7.3)$ km/s {\bf (Sch{\"o}nrich} 
et al., 2010) is the solar motion with respect to LSR , and

\[\left[\begin{array}{lll}
 & U & \\
 & V & \\
 & W & 
\end{array} \right] = B\left[\begin{array}{lll}
  & k\mu_{\alpha }/\pi   & \\
 & k\mu_{\delta }/\pi  & \\
& \rho  & 
\end{array} \right]\]
where B = T.A, k = 4.74057 km s$^{-1}$, $\mu _{\alpha }$ is the proper motion 
in right ascension, in arcsec yr$^{-1}$, $\mu _{\delta }$ is the 
proper motion in declination, in arcsec yr$^{-1}$, $\rho$ is the 
radial velocity in kms$^{-1}$  and $\pi$ is the parallax in arcsec.

The transformation matrix connecting 
Galactic coordinates and equatorial coordinates is:
 \[T = \left[\begin{array}{ccc}
-0.0548756 & -0.8734371 & -0.4838350 \\
+0.4941094  & -0.4448296 & +0.7469822  \\
-0.8676661 & -0.1980764 & +0.4559838 
\end{array} \right]\]
Coordinate matrix A can be defined as:\\

\[A = \left[\begin{array}{ccc}
$-$sin\alpha  & $-$cos\alpha sin\delta  & +cos\alpha cos\delta  \\
+cos\alpha    & $-$sin\alpha sin\delta  & sin\alpha cos\delta   \\
0  & +cos\delta & sin\delta 
\end{array} \right]\]
 Where $\alpha$ is the right ascension and $\delta$ is the 
declination in degrees. We are using a right-handed coordinate 
system for U,V and W. Hence they are positive in the directions 
of the Galactic center, Galactic rotation, and the North Galactic 
Pole (NGP) respectively (Johnson \&  Soderblom (1987)).

The estimates for
proper motion are taken from SIMBAD. Distances are measured taking
parallax values from SIMBAD and \textit{Gaia} (Gaia Collaboration 2016, 
Gaia Collaboration et al. 2018b)
whenever available. We have used the  radial velocity estimates and the
corresponding error estimates as  measured by us. 

\noindent
The total spatial velocity of a star is given by,\\
  $V_{spa}^{2}=U_{LSR}^{2}+V_{LSR}^{2}+W_{LSR}^{2}$

The probability for a star's  membership into the thin disk, the 
thick disk or the halo population are also calculated 
following  the procedures of Reddy, Lambert \& Priesto (2006), Bensby et al. (2003, 2004),
Mishenina  et al. (2004), and presented in Table 7, along with the 
estimated components of spatial velocity and the total spatial
velocity.
The selection of the thick or thin disk stars and halo stars are 
based on the assumption that the Galactic space velocities of 
these stars have Gaussian distributions.\\
$f(U,V,W) = K\times\exp[-\frac{U_{LSR}^{2}}{2\sigma _{U}^{2}}-\frac{(V_{LSR}-V_{asy})^{2}}{2\sigma _{V}^{2}}-\frac{W_{LSR}^{2}}{2\sigma _{W}^{2}}]$\\
where\\
$K = \frac{1}{(2\pi) ^\frac{3}{2}\sigma _{U}\sigma _{V}\sigma _{W}}$

The values of the characteristic velocity dispersion $\sigma _{U}$, 
$\sigma _{V}$  and $\sigma _{W}$   and the asymmetric drift
$V_{asy}$   are  adopted from Reddy et al. (2006).

{\footnotesize
\begin{table*}
{\bf Table 7: Spatial velocity and probability estimates for 
the programme stars}\\ 
\begin{tabular}{lcccccccc}
\hline                       
Star name           & U$_{LSR}$         & V$_{LSR}$           & W$_{LSR}$ & V$_{spa}$  & p$_{thin}$ & p$_{thick}$ & p$_{halo}$ & Population \\
                    & (kms$^{-1}$)        & (kms$^{-1}$)          & (kms$^{-1}$) &  (kms$^{-1}$) &           &           & &  \\  
    \hline
    HE~0308$-$1612  &  $-180.36$$\pm$17.11& $-90.63$$\pm$10.0 &$49.69$$\pm$13.22& $207.87$$\pm$15.98& 0.0 &0.85   & 0.15 & Thick disk  \\
    CD$-28$ 1082    & $-3.95$$\pm$1.17  & $-132.96$$\pm$6.99 & 107.91$\pm$3.53  & 171.28$\pm$3.23 & 0.0  & 0.73  & 0.26 & Thick disk  \\
    HD 29370        & 22.08$\pm$0.29    & 18.85$\pm$0.19     & $-$1.95$\pm$0.24 & 29.09$\pm$0.33  & 0.99  & 0.01  & 0.0 & Thin disk\\
    HD 30443        & $-$60.89$\pm$0.25 & $-2.20$$\pm$0.62    & 11.72$\pm$0.30   & 62.04$\pm$0.21  & 0.98  & 0.02  &0.0 & Thin disk \\
    CD$-38$ 2151    & $-$68.49$\pm$1.25 & $-126.23$$\pm$1.43 & 6.64$\pm$3.01    & 143.76$\pm$1.71 & 0.02  & 0.95  & 0.03 & Thick disk  \\
    HD 50264        & $-48.07$$\pm$1.35 & $-34.99$$\pm$2.19  & 11.67$\pm$0.68   & 60.59$\pm$2.21  & 0.97  & 0.03  & 0.0 & Thin disk \\
    HD 87080      & $-87.39$$\pm$1.39   & 16.13$\pm$0.14     & 23.55$\pm$0.23   &  91.94$\pm$1.24 & 0.95  & 0.05  & 0.0 & Thin disk \\
    HD 87853     & $-$43.92$\pm$0.96   & 21.24$\pm$0.24     & $-$24.11$\pm$0.50& 54.42$\pm$0.90  & 0.98  & 0.02  & 0.0 & Thin disk\\
    HD 123701     & 4.71$\pm$0.19       & 30.51$\pm$0.24     & $-0.88$$\pm$0.19 & 30.88$\pm$0.26  & 0.99 & 0.01 & 0.00 & Thin disk\\           
    HD 176021     &  75.65$\pm$0.41     & $-113.05$$\pm$0.33 & 0.05$\pm$0.28    & 136.03$\pm$0.05 & 0.07  & 0.90  & 0.02 & Thick disk   \\
    HD 188985     &  26.74$\pm$0.05     & 17.18$\pm$0.03     & 18.51$\pm$0.07   & 36.78$\pm$0.09  & 0.99 & 0.013  & 0.0 & Thin disk \\
    HD 202020     & 6.56$\pm$0.11       & 10.98$\pm$0.11     & 33.28$\pm$0.16   & 35.65$\pm$0.21  & 0.96  & 0.034  & 0.0 & Thin disk\\
\hline
\end{tabular}
\end{table*}
}

\section{DISCUSSION}

In the following, we discuss  the abundance patterns and other
properties of the individual stars.

\par {\bf HE 0308-1612 -}
This object is listed  among the 403 faint high latitude carbon stars 
of Hamburg/ESO survey(HES; Christlieb et al. 2001). From low-resolution 
spectroscopic analysis Goswami (2005) confirmed  this object to be  a 
potential CH star. In this work we have conducted a detailed chemical  
analysis  based on a high resolution spectrum. The object is found to 
be  mildly metal-poor  with a metallicity of $-$0.73. Further, with 
its estimated  [C/Fe]$\sim$0.78, [Ba/Fe]$\sim$1.63  and [Ba/Eu]$\sim$1.28, 
the star may be placed in the group of CEMP-s stars. As the oxygen  
abundance could not be estimated, this object could not be classified 
on the basis of C/O ratio. The star has a low value of 15.6 for 
$^{12}$C/$^{13}$C ratio.

\par {\bf CD$-$28 1082 -} 
This star belongs to the CH star catalogue of Bartkevicius (1996).
This work presents the first time abundance analysis  for this object. 
The object  is found to be a very metal-poor  with a metallicity of $-$2.45.
Carbon is highly enhanced  with [C/Fe] $\sim$ 2.19. Barium as well 
as europium are also found to be enhanced relative to iron. The
star satisfies the classification  criteria for CEMP-r/s stars
([C/Fe] $>$ 1, [Ba/Fe] $>$ 1 and 0.0 $\leq$ [Ba/Eu] $\leq$ 0.50) of  
Beers \& Christlieb (2005)). 

\par A comparison of the  the abundance ratios  of light 
and heavy elements in CD$-28$ 1082 with those  observed in  
CEMP-r/s and CH stars from literature  (Figs 
18 and 19) shows a good match with  the CEMP-r/s 
stars.  Three CEMP-r/s stars from Aoki et al.(2007) exhibit very 
high  Mg abundances  (CS 29497-034 ([Mg/Fe] = 1.31),
CS 29528-028 ([Mg/Fe] = 1.69), and HE 1447+0102 ([Mg/Fe] = 1.43)).
 The astrophysical processes that could produce  such high enhancement of
Mg are  not  clearly understood (Aoki et al. 2007).  Our estimated 
value of  [Mg/Fe] (= 0.45)
for CD$-$ 28 1082 is close to that  found by Magain (1989), who had
obtained a  mean value of [Mg/Fe] = 0.48$\pm$0.09 dex from an analysis
of 20 extremely metal-poor
halo dwarfs in the  metallicity range $-2.9$ $<$ [Fe/H] $<$ $-1.4$.

Analysis of  extremely metal-poor field giants in the
metallicity range $-3.0$ $<$ [Fe/H] $<$ $-1.3$ by 
Gratton \& Sneden (1988) suggested
a mean value of [Mg/Fe] = 0.27 dex with a star-to-star scatter
of 0.14 dex.    [Mg/Fe] is found to be increasing with decreasing
[Fe/H].  The analysis of  Edvardsson
et al. (1993)  shows  [Mg/Fe] $\sim$ 0.5 dex at [Fe/H] = $-1.0$ dex
and attains near solar value at [Fe/H] = 0.0 dex. 
They also report anomalously large value of [Mg/Fe] for a few
of the disk dwarfs in their sample.  de Castro et al. (2016) 
analysed the variation of [Mg/Fe] with [Fe/H] for both barium
and field giants that  lie in a range from
 0.0 to 0.5-0.6 at [Fe/H] $\sim$ $-1.5$. In the  
log($^{12}$C/$^{13}$C) versus [C/N] plot 
of  Masseron (2010), this object falls well within  the region 
occupied by CEMP stars (Fig 20). 

\par From kinematic analysis,  CD$-28$ 1082 is found to be 
a thick  disk star with a probability of nearly 73 percent. The 
estimated spatial velocity (${\sim}$ 180 km s$^{-1}$) is similar 
to that expected for  halo objects (Chen, Nissen \& Zhao 2004). This object also 
satisfies the condition for  V$_{LSR}$ $<$ -120Km s$^{-1}$ for it 
to be a halo star (Eggen 1997). 
\begin{figure}
\centering
\includegraphics[width=\columnwidth]{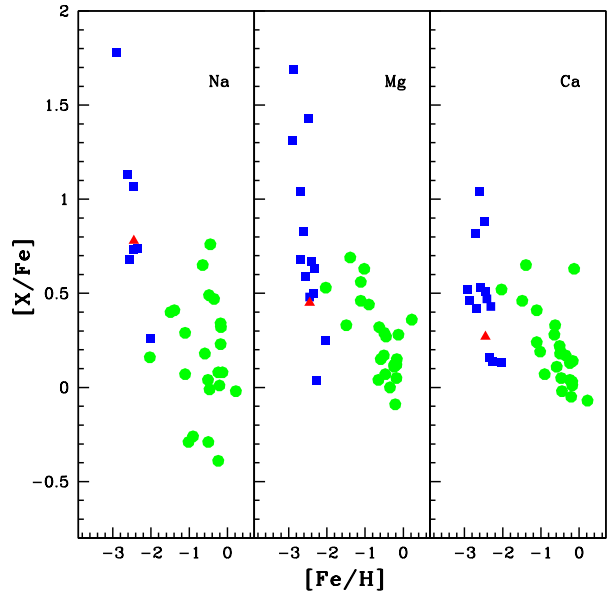}
\caption{Comparison of abundances of light elements for CH stars 
(Sneden \& Bond 1976; Luck \& Bond 1991; Vanture 1992; Karinkuzi \& Goswami 2014, 2015)
 and CEMP-r/s stars (Masseron et al. 2010). Filled circles represent CH stars and the filled squares 
represent CEMP-r/s stars. The object CD$-28$ 1082 is represented
by a  filled triangle.}
\end{figure}

\begin{figure}
\centering
\includegraphics[width=\columnwidth]{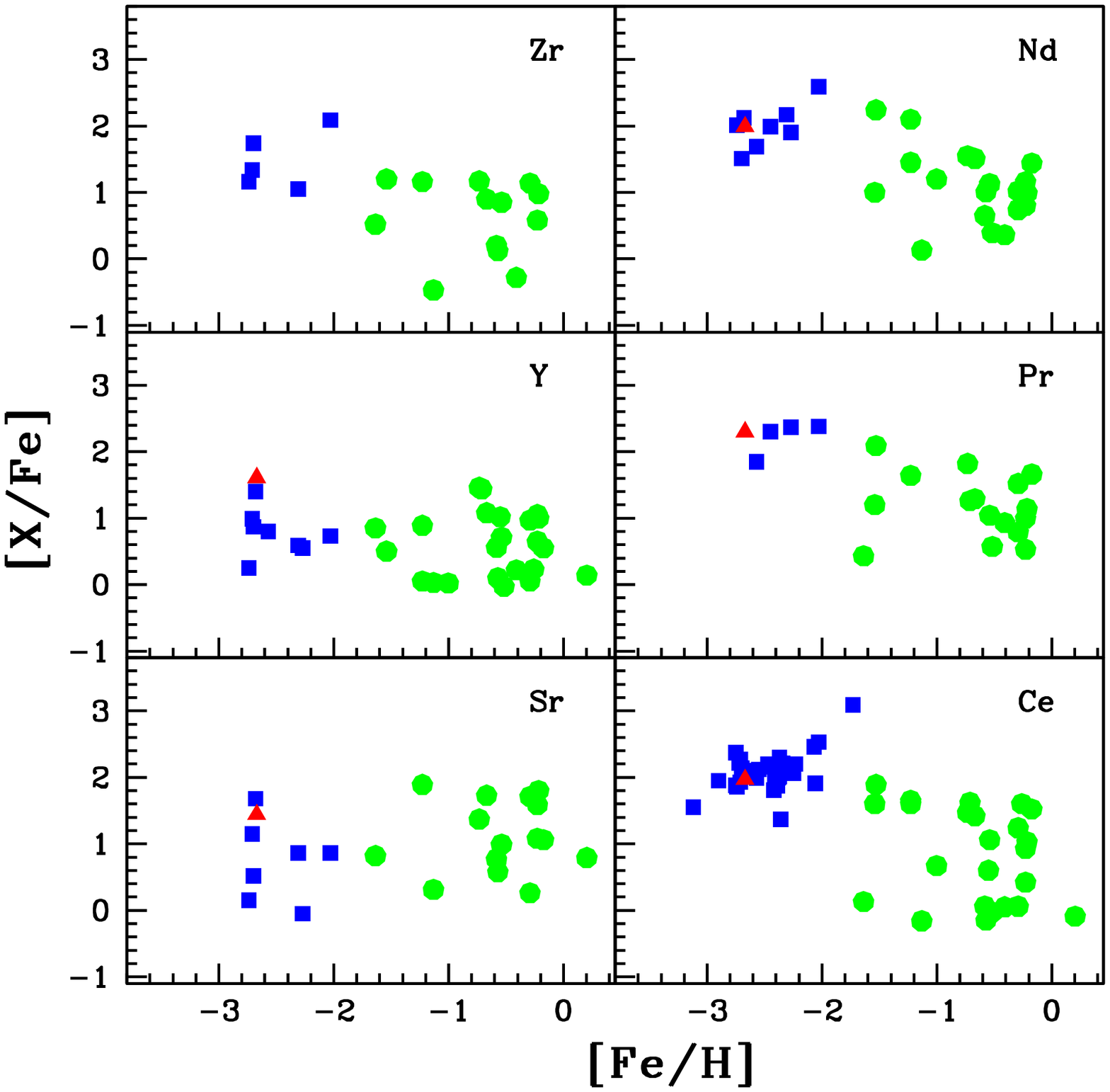}
\includegraphics[width=\columnwidth]{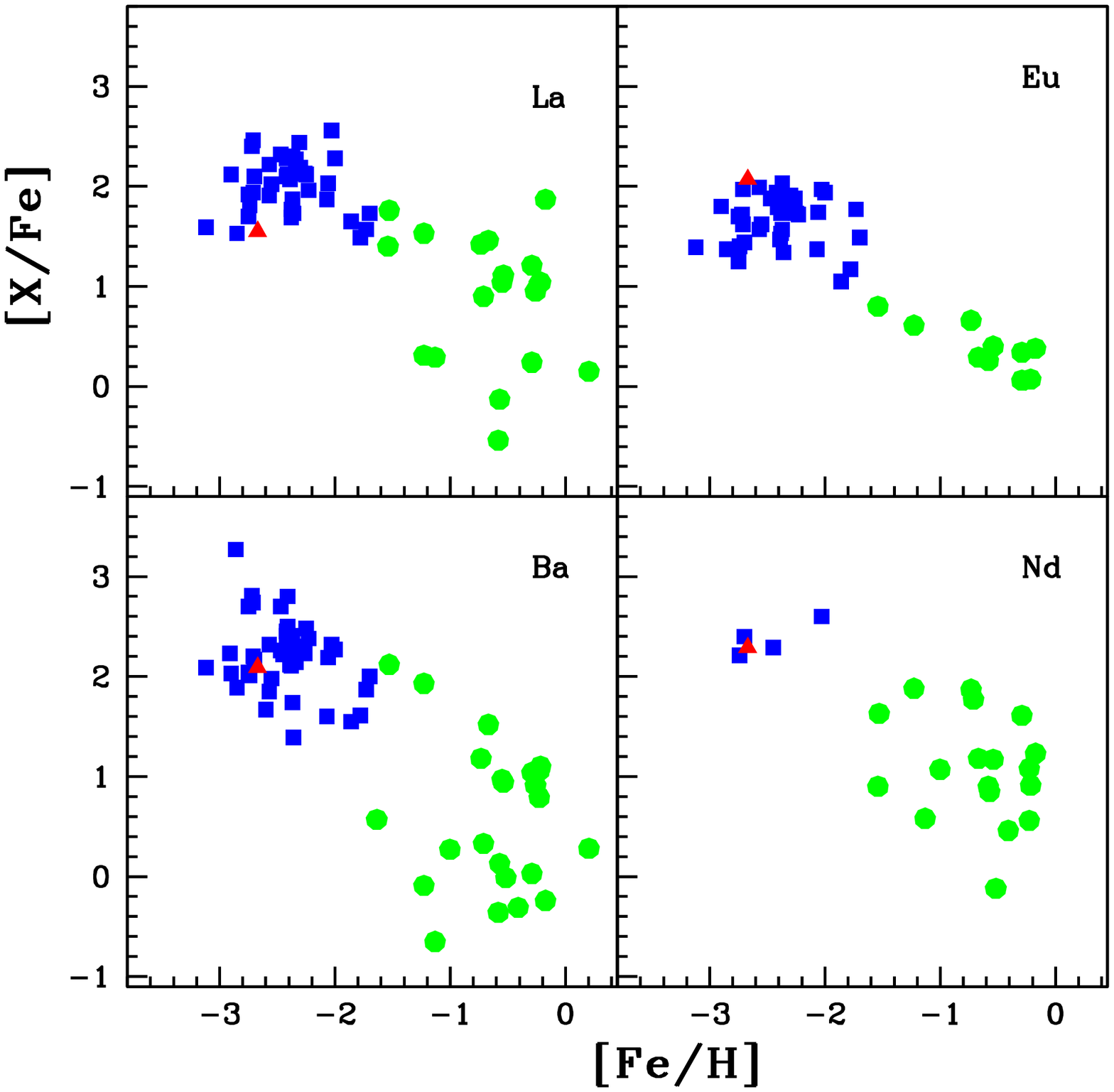}
\caption{Comparison of abundances of heavy elements for CH stars 
(Sneden \& Bond 1976; Luck \& Bond 1991; Vanture 1992; Karinkuzi \& Goswami 2014, 2015) and CEMP-r/s stars 
(Masseron et al. (2010)). Filled circles represent
CH stars and the  filled squares represent CEMP-r/s stars. 
The object CD$-28$ 1082 is represented by a  filled triangle.
}
\end{figure}

\par  There exist several physical scenarios on the formation
of CEMP-r/s stars (Jonsell et al. 2006, Bisterzo et al. 2011 and 
references there in).  One of the possible scenarios 
is that, the CEMP-r/s stars  are formed in a medium enriched 
in r-process elements which is produced by a  supernovae 
explosion in the past.  Argast et al. (2004)
studied the interstellar enrichment of r-process elements in the
context of inhomegenous chemical evolution. Their studies show that
core collapse  supernovae are the dominant source of r-process
elements compared to neutron star mergers. Calculations of
Rosswog et al. (1999, 2000) show that although the amount of r-process
matter ejected in neutron star merger is much larger than that
in supernovae collapse, the rate of neutron star merger in
the galaxy is less than the Type II SNe (Tammann, L$\ddot{o}$ffler \& Schr$\ddot{o}$der 1994;
Belczynski, Kalogera \& Bulik 2002 ). Hence it is likely that both of these 
sources might  contribute
to the r-process elements in the Galaxy (Qian \& Woosley 1996;
Rosswog et al. 1999; Rosswog \& Davies 2002; Thielemann et al. 2002).

Bisterzo et al. (2011) suggest that the molecular cloud from which
the binary system formed was previously polluted in r-process 
elements by some r-process sources,  believed to have occurred
during supernovae explosions and /or neutron star mergers (Thielemann 
et al. 2011, Wehmeyer et al. 2015). Argast et al. (2004) rule out
neutron star merger as a dominant source of r-process
since the expected r-process enrichment by neutron star merger
is not found to be consistent with the observations at very low 
metallicity stars. 
 However, Van de Voort et al. (2015) studied the abundance pattern of 
r-process elements using cosmological zoom-in simulations of a Milky 
Way-mass galaxy using  various models for the rate and delay time of 
neutron star mergers and obtained  results  that are in contrast to those  
of Argast et al. (2004). 

\par It was shown in Van de Voort et al. (2015) that the neutron star mergers 
can produce [r-process/Fe] abundance ratios and scatter, 
consistent with the observations for stars with metallicity 
in the range $-2.0$ $\leq$ [Fe/H] $\leq$ 0. Despite  some 
uncertainties in their model calculations, it was concluded that  
neutron star mergers can be an important source of the majority of 
the r-process elements. Investigation of the production sites of 
r-process elements in the Galaxy by  Shen et al. (2015) based on the 
high resolution cosmological zoom-in simulation also concludes that neutron 
star mergers can be a dominant production site of r-process elements even 
at low metallicity. Haynes et al. (2018) studied the distribution of 
elemental abundance ratios using chemodynamical simulations including 
various neutron capture  processes such as magneto-rotational supernovae, 
neutron star mergers, neutrino driven winds and electron capture supernovae. 
A comparison of their  predicted trend between [Eu/Fe] and [Eu/O] as 
a function of metallicity  with the observational data shows that  
neither electron capture supernovae nor neutrino driven  winds could 
explain the observed Eu levels; however, neutron star mergers and 
magneto-rotational supernovae are able to justify the observed Eu abundances. 
\par The possible explantion for the observed 
s-process enhancemet is attributed to a binary companion. 
Another scenario of CEMP-r/s stars formation  is that it is born in 
a triple system (Cohen et al.  2003), in which one of the companions 
explodes as supernova. The tertiary star accretes the 
r-process rich material from this companion. The 
enhancement in carbon and s-process elements are explained 
by the accretion of materials from its secondary AGB 
companion. Cohen et al. (2003) also suggested another mechanism 
in which  the CEMP-r/s stars are initially polluted by its binary 
companion which later becomes a white dwarf. At some point 
the reverse mass transfer can take place from the secondary 
star to the white dwarf. This may induce accretion induced 
collapse of the white dwarf if it is an O-Ne-Mg dwarf. 
The resultant neutrino driven wind is sufficient to contribute 
r-process elements on the surface of the secondary star. 
Yet another possible scenario is that the 
CEMP-r/s star is in a binary system and accretes material 
from its primary companion. Later the companion explodes 
as a  type 1.5 supernova (Zijlstra, 2004). Low metallicity 
stars with initial mass 3-4 M$\odot$ undergo this type of 
evolution (Zijlstra 2004). Abate, Stancliffe \& Liu (2016) have examined
the possibility of different mechanisms of formation of 
CEMP-r/s stars but  could not reproduce the observed frequency 
of CEMP-r/s stars by their simulations.

\par  Several authors (Hansen et al. 2016; Hampel et al. 
2016 and references therein)  have discussed the 
possibilities of i-process in  low metallicity AGB 
stars as the origin of CEMP-r/s  stars. The neutron 
density associated with the i-process  is sufficient to 
produce both the s and r process elements in AGB stars. 
Hampel et al. (2016) showed that the observed abundance pattern 
in many of the CEMP-r/s is consistent with the model 
calculations for i-process within the observational errors. 
It would be worthwhile to check  in the future if 
 the abundance pattern observed in CD$-$28 1082
can be explained on the basis of  the i-process.

\par {\bf CD$-$38 2151 - }
 This star belongs to the CH star catalogue of Bartkevicius (1996).
Our  abundance estimates  are also  characteristic of giant CH
stars. The probability estimate shows this to be a thick disk object.
 With the lower bound, the spatial velocity estimate 
supports the inclusion of the star into the disk population while 
the upper bound in spatial velocity and the metallicity estimate 
shows this to be a halo object. It also satisfies the condition 
V$_{LSR}$ $<$ $-$120 Kms$^{-1}$ for it to be a halo star  (Eggen 1997). 

Previous studies on this object include an  abundance analysis  
by Vanture (1992) who  had  performed abundance estimates 
for six heavy elements based on spectra with typical 
resolution of ${\lambda}/{\delta\lambda}$ ${\sim}$ 20000. 
Our detailed abundance analysis for both heavy and  light 
elements are based on higher resolution spectra. The star is found 
to be  very metal-poor  with a metallicity  (${\sim}$  $-$2.0). 
Carbon is enhanced with [C/Fe] $\sim$ 1.50. The heavy s-process 
element barium is  also enhanced. However, as we could not estimate the 
 Eu abundance, its membership for CEMP-r/s class of objects
could not be examined  based on the CEMP stars classification 
scheme of  Beers \& Christlieb (2005). A comparison of our elemental
abundance ratios  with literature values including  Vanture's are 
presented in Table 8. Although our T$_{eff}$ estimate is close to
 Vanture's,  log\,{g} value differs from his. While for La and 
Ce our estimates are lower than Vanture's, for Nd and Sm
our estimates are higher. 
 Spectrum synthesis calculation of La II  4921.77 \AA\,  
gives a better fit  with our adopted La abundance
 than that obtained  using Vanture's atmospheric 
parameters and abundance value of La.   
In the log($^{12}{C}$/$^{13}{C}$) versus [C/N] plot (Fig. 20),
CD$-38$ 2151 falls in the region occupied by the CEMP stars.
Estimated C/O ${\sim}$ 1.26 is similar
to values seen in CH stars.

\begin{figure}
\centering
\includegraphics[width=\columnwidth]{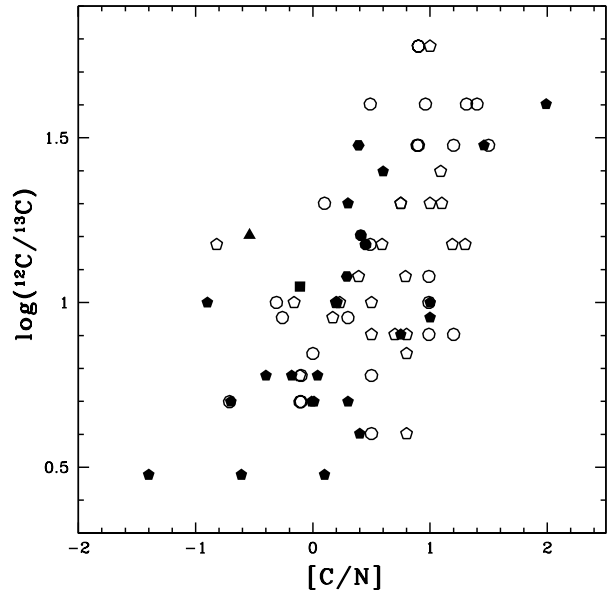}
\caption{log($^{12}{C}$/$^{13}{C}$) versus [C/N] for CEMP stars 
is presented. The data are taken from Masseron et al. (2010).
Open circles represent CEMP-s stars, open pentagon corresponds to 
CEMP-r/s stars and CEMP-r stars are represented by filled 
circle. Filled pentagon corresponds to CEMP-no stars and 
CEMP-low s stars are represented by filled hexagon. 
CD-28 1082 is represented by a filledd 
triangle  and  CD-38 2151 by a filled square. }
\end{figure}   

\par  {\bf HD 30443 - } This star belongs to the CH star catalogue of
 Bartkevicius (1996). Yamashita (1975) also had classified this
object as a CH star. This work presents  for the first time a 
detailed abundance  analysis for this object. The estimated  metallicity 
is   $-1.68$. The star is enhanced in Carbon with [C/Fe]$\sim$1.68 
as well as in other heavy elements. The abundance estimates of 
the star are  characteristic of CH giants. The estimated 
C/O $\sim$ 1.02 is similar to values seen in CH stars. 
From the kinematic analysis,  HD 30443 is found to be a 
thin disk star with a probability of  98 percent. The estimated 
spatial velocity is $\sim$ 62.04 km $s^{-1}$, which is similar 
to that expected for thin disk objects (Chen et al. 2004). 

\par {\bf HD~29370 and HD~123701 - } 
These two  objects are  listed in both the CH star catalogue of 
Bartkevicius (1996) and the barium star catalogue of L\"u (1991). 
Sleivyte \&
Bartkevicius (1990) based on photometric analysis included 
 HD~29370 in  the category of metal-deficient 
barium stars.  However,  Catchpole, Robertson \& Warren (1977) classified 
HD~29370 as a metal weak CH star. Mennessier et al. (1997) 
suggested this object to be either a subgiant or a giant star. 
From our estimates of luminosity and  temperature,  we have
confirmed it to be a giant star from its location in the 
H-R diagram.  de Castro et al. (2016) have 
estimated the abundance of the Na, Al, $\alpha$-elements, 
iron-peak elements, and s-process elements Y, Zr, La, Ce, and 
Nd for these two objects. In addition to these elements 
we have derived abundances 
for elements  C, N and O for HD~123701  and C and N for 
HD~29370 which were not presented in de Castro et al. (2016). 
Our analysis confirms the object HD~123701 to be a barium star
with C/O ${\sim}$ 0.15. As the oxygen abundance could not be estimated
for HD~29370, it could not be classified based  on estimates of
the C/O ratio.

\par {\bf HD~50264 -} This object is listed in both the CH star catalogue of 
Bartkevicius (1996) and the barium star catalogue of L\"u (1991). 
While  Bond (1974)  considered this star as a CH subgiant,
 Mennessier et al. (1997) grouped HD 50264 as a dwarf. According to
the analysis of Pereira \& Junqueira (2003) both HD 50264 and HD 87080 
are CH subgiants.  
Our analysis shows the object to exhibit  characteristic properties
of barium stars with C/O ratio ${\sim}$ 0.41.

\par {\bf HD~87080 -} 
This object is listed in both the CH star catalogue of
Bartkevicius (1996) and the barium star catalogue of L\"u (1991). 
Mennessier et al. (1997)  assigned HD~87080 to group of 
subgiants. Our work  confirms the same. 
A comparison of our estimates of elemental abundance ratios  
with those of Allen \&
Barbuy (2006a,b) is presented in Table 8. For most of the elements 
we find a good agreement.  This 
object  shows the characteristic properties of barium stars with 
C/O ratio ${\sim}$ 0.29. 

\par {\bf HD 87853 - }
This star is found to be mildly metal-poor with an estimated 
metallicity of $-$0.73. All the elements show  abundances 
as expected for  normal 
metal-poor star of similar metallicity. This is true for 
$\alpha$-elements  as well as for heavy neutron capture elements. 
The position of the star on the HR diagram indicates that the 
star has just left the  main-sequence, approaching the subgiant 
stage. 

\par {\bf HD~176021 -} 
Bond (1970)  considered this  star as a subgiant CH star. 
This star belongs to the CH star catalogue of Bartkevicius (1996).
While Bensby, Feltzing \& Oey (2014) derived  abundances  for a
few  elements,  Battistini \& Bensby (2015) 
determined the abundances  of Sc, V, Mn and Co
to study the origin and evolution  of these elements in 
a homogenous sample of stars.  According to our abundance 
estimates, the star qualifies to be a sub-giant CH star. We 
have also measured abundances of carbon and nitrogen,
and both are found to be similarly enhanced. The ratio 
of C/O could  not be estimated for this object. 

\par {\bf HD~188985 -} 
This object is listed in both the CH star catalogue of
Bartkevicius (1996) and the barium star catalogue of L\"u (1991). 
North et  al. (1994) and Mennessier et al. (1997) placed this
object  as a barium dwarf. Our analysis confirms the same.  
Our abundance estimates match closely with those of Allen \&
Barbuy (2006a,b). Estimated C/O ratio
for this object is ${\sim}$ 0.89.

\par {\bf HD 202020 -} 
From radial velocity monitoring, McClure (1997) confirmed 
HD 202020 to be a  binary  with an orbital period of 2064$\pm 10$ d.
 The first abundance analysis for this star was performed  by Luck 
\& Bond (1991); however abundances of
elements such as carbon, Nitrogen, Strontium and Barium were not
presented. We have estimated the abundances of these elements 
in the present work. HD~202020 is listed in  Bartkevicius (1996).
There is a discrepancy regarding the oxygen abundance in this star.
If we consider the value of oxygen abundance obtained using  the line 
at 6300.3 {\rm \AA}, estimated C/O ratio turns out to be  $>$ 1, 
a value generally seen in case of  CH stars. With C/O ${\sim}$ 1.99, the
star qualifies to be a CH star. We get a value much  lower than one,
 if we consider the non-LTE corrected abundance value  of oxygen obtained 
from the oxygen triplet lines. We note that, a  value 
of C/O $<$ 1 is generally seen for barium stars.

\par  The position of our programme stars in the 
colour-magnitude diagram is shown in Fig. 21. A comparison of 
 the abundances of both light and heavy elements of 
our programme stars with those  of Ba stars, CH stars 
and CEMP stars from literature is   shown 
in Figs 22 and 23, respectively.

\begin{figure}
\centering
\includegraphics[width=\columnwidth]{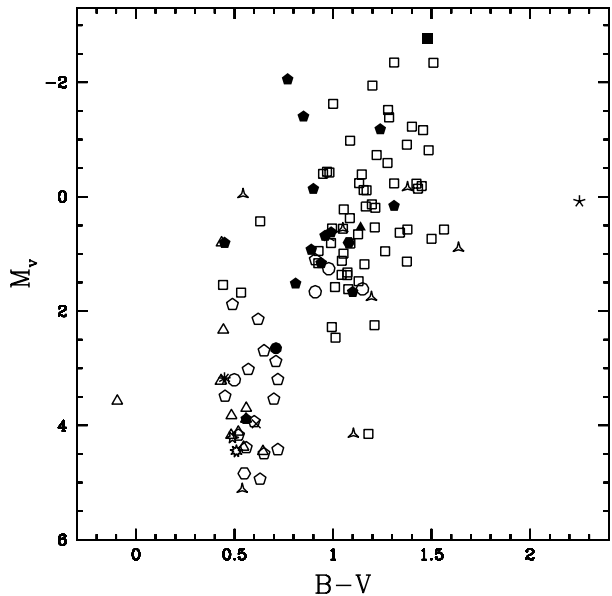}
\caption{ The locations of our programme stars, CD$-28$ 1082 (filled triangle),
 CD$-38$ 2151 (filled square), HD  50264 (Open hexagon), HD 87080 (filled 
circle), HD 123701 (skelton triangle), HD 176021 (cross), HD 188985 (star), 
HD 29370 (filled hexagon),  HD 30443 (five sided cross), 
HE 0308$-$1612 (six sided cross), HD~87853 (nine sided cross), and  
HD 202020 (nine sided star),  in the colour-magnitude diagram. 
Open squares represent Ba giants, open triangle is for Ba dwarfs and open 
circle represents barium subgiants from literature (de Castro et al. 2016; 
Yang et al. 2016; Allen \& Barbuy 2006). CH subgiants are 
represented by open 
pentagon and filled pentagon represents CH giants from literatures 
(Karinkuzhi \& Goswami 2014,2015; Goswami et al. 2016). CEMP stars are 
represented by starred triangle (Masseron et al. 2010). }
\end{figure}

\begin{figure}
\centering
\includegraphics[width=\columnwidth]{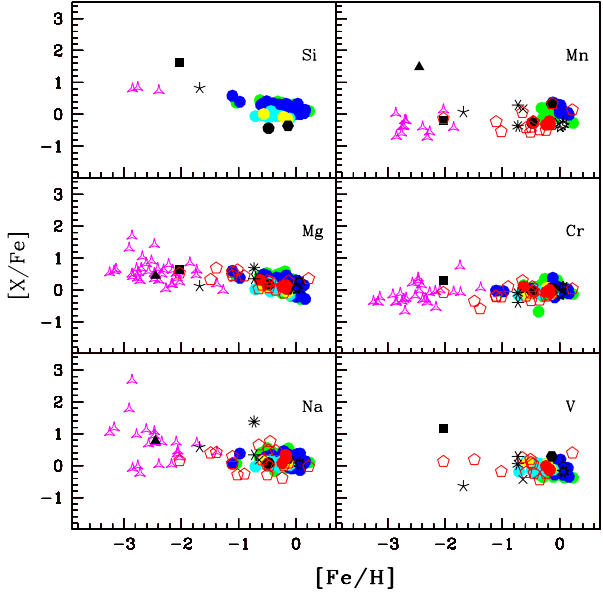}
\includegraphics[width=\columnwidth]{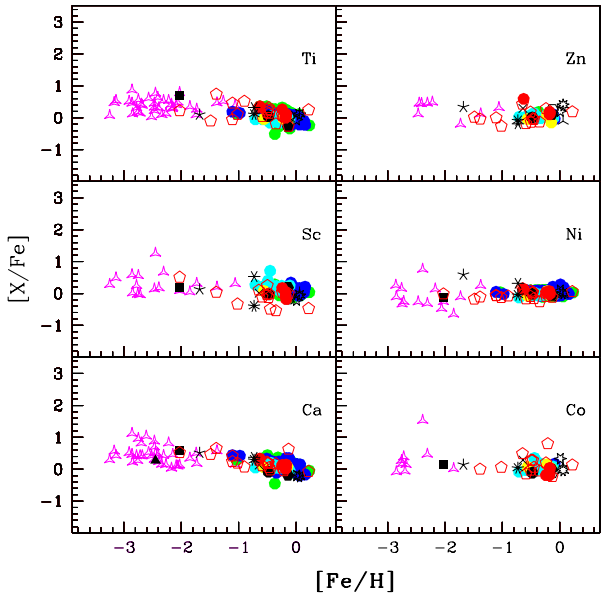}
\caption{ Abundance ratios of light elements with respect to metallicity 
[Fe/H] observed in our programme stars are represented in different symbols. 
CD$-28$ 1082 (filled triangle), CD$-38$ 2151 (filled square), HD  50264 
(Open hexagon), HD 87080 (filled circle), HD 123701 (skelton triangle), 
HD 176021 (cross), HD 188985 (star), HD 29370 (filled hexagon),
HD 202020 (nine sided star), HD~30443 (five sided cross), HE~0308$-$1612 
(six sided cross) and HD~87853 (nine sided cross). Green filled circles
represent strong Ba giants, 
blue filled circles represent weak Ba giants, cyan filled circles are
 for Ba dwarfs 
and yellow filled circles represent barium subgiants from literature 
(de Castro et al. 2016; Yang et al. 2016; Allen \& Barbuy 2006). CH subgiants are 
represented by red filled circles, and  red open  pentagons represent
 CH giants from  literature (Karinkuzhi \& Goswami 2014,2015; Goswami 
et al. 2016). CEMP stars are represented by magenta starred 
triangle (Masseron et al. 2010).  }
\end{figure}

\begin{figure}
\centering
\includegraphics[width=\columnwidth]{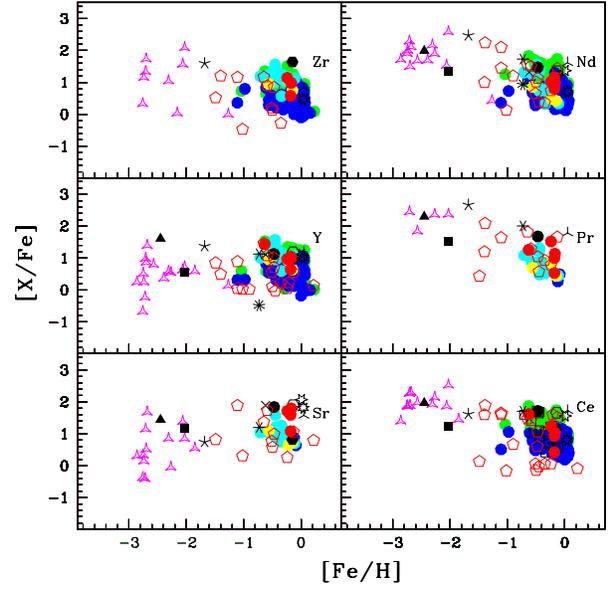}
\includegraphics[width=\columnwidth]{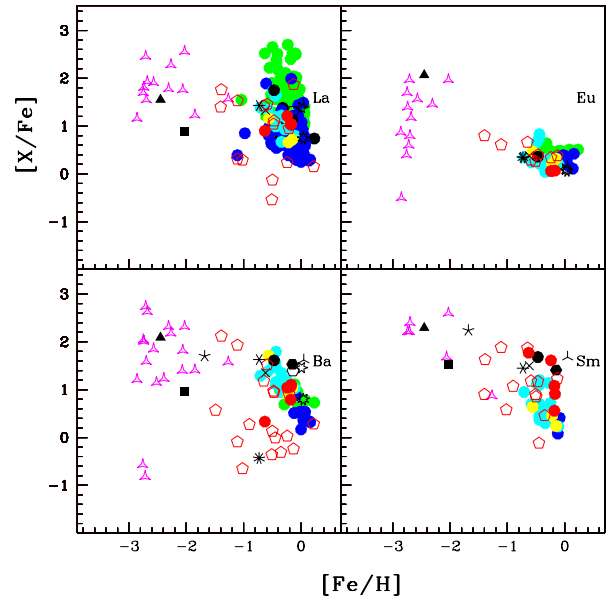}
\caption{  Same as Figure 22 for heavy elements.}
\end{figure}

{\footnotesize
\begin{table*}
{\bf Table 8 : Comparison of the abundances of our programme stars with the literature values. }\\ 
\begin{tabular}{lcccccccc}
\hline                       
Star name & [Fe I/H] & [Fe II/H] & [Fe/H] & [Sr/Fe] & [Y/Fe] & [Zr/Fe] & [Ba II/Fe] &  Ref \\
\hline
HE~0308$-$1612 & $-$0.72 &$-$0.73 &$-$0.73 & 1.20    & 1.13   & --     & 1.63 & 1   \\
CD$-28$ 1082   & $-2.46$ & $-2.44$  & $-2.45$   & 1.44 & 1.61 &  - & 2.09   & 1\\
HD 29370       & $-0.15$ & $-0.13$ & $-0.14$ & 0.82 &0.87 & 1.64 & 1.53 & 1 \\
               & $-0.25$ & $-0.28$ & $-0.27$ &  -   & 0.83 & 0.71 & - & 2 \\ 
HD 30443       & $-$1.68 & $-$1.69 &$-$1.68  & 0.74 & 1.37 & 1.60 & 1.70 & 1 \\
CD$-38$ 2151   & $-2.03$ & $-2.03$ & $-2.03$ & 1.16 & 0.57 & 2.00  & 0.97 & 1 \\
               &    -    &   -     &  $-1.40$&   -  & 0.30 &   -     & - & 3 \\
HD 50264       & $-0.14$ & $-0.11$ & $-0.13$ & 1.90 & 1.16 &  -      & 1.40 & 1 \\
               & $-$0.34 & $-$0.37 & $-0.34$ & -    & 0.88 & 1.29    & 1.25 & 4 \\
HD 87080       & $-0.47$ & $-0.49$ & $-0.48$ & 1.84 & 1.12 &  -      & 1.61 & 1 \\
               & $-$0.51 & $-$0.49 & $-0.51$ & -    & 1.01 & 1.22    & 1.51 & 4 \\
               & $-0.49$ & $-0.44$ & -       & 0.98 & 1.11 & 1.35    & 1.48 & 5\\
HD 87853       & $-$0.72 &$-$0.74  & $-$0.73 & --   & $-$0.48 & --   & $-$0.43 & 1 \\ 
HD 123701      & 0.05    & 0.07    & 0.06    & 1.58 & 1.17 & 0.83    & 1.62 & 1 \\
               & $-0.44$ & $-0.46$ & $-0.45$ &  -   & 0.72 & 0.94    & -    & 2\\
HD 176021      & $-0.62$ & $-0.65$ & $-0.64$ & 1.89 & 1.11 &  -      & 1.35 & 1 \\
               & -       & -       & $-0.40$ & 0.95 & 0.75 & 0.44    & 1.22 & 6 \\
HD 188985      & $ 0.01$ & $ 0.01$ & $ 0.01$ & 2.07 & 1.13 &  -      & 1.46 & 1 \\
               & $-0.25$ & $-0.30$ & -       & 1.05 & 1.02 & 1.18    & 1.20 & 5 \\  
               & $-0.51$ & $-0.32$ & -       & 1.10 & 0.95 & 0.85    & 1.27 & 7\\
HD 202020      & 0.04    & 0.08    & 0.06    & 1.82 & 1.00 & 0.48    & 0.80 & 1\\
               & $-$0.14 & $-$0.07 & $-0.13$ &  -   & 0.80 & 0.25    &   -  & 8 \\\\
                       
Star name    & [La II/Fe] & [Ce II/Fe] & [Pr II/Fe] & [NdII/Fe] & [Sm II/Fe] & [Eu II/Fe] &  & Ref \\\\

HE 0308$-$1612 & 1.43     & 1.71       &2.01        & 1.73      & 1.45       & 0.35       &           & 1 \\
CD$-28$ 1082 & 1.55       & 1.97       & 2.30      & 1.99      & 2.29       & 2.07        &           & 1 \\ 
HD 29370     & 1.13       & 1.14       & -          & 1.19      & 1.41       &  -         &           & 1 \\
             & 1.61       & 1.03       & -          & 1.00      &  -         &  -         &           & 2 \\
HD 30443     &  --        & 1.62       & 2.67       & 2.47      & 2.24       & -          &           & 1 \\
CD$-38$ 2151 & 0.88       & 1.24       & 1.52       & 1.34      & 1.53       &  -         &           & 1 \\
             & 1.30       & 1.35       & -          & 1.00      & 0.61       &  -         &           & 3 \\
HD 50264     & 1.30       & 1.66       & -          & 1.35      & -          &  -         &           & 1\\
             & 0.92       & 0.76       & -          & 0.47      & -          & 0.44       &           & 4 \\
HD 87080     & 1.75       & 1.71       & 1.68       & 1.47      & 1.68       & 0.36       &           & 1 \\
             & 1.75       & 1.32       & -          & 0.97      & -          & 0.61       &           & 4 \\
             & 1.74       & 1.73       & 1.24       & 1.56      & 1.12       & 0.66       &           & 5 \\
HD 87853     & -          & -          &  -         & 0.94      & -          & 0.34       &           & 1  \\
HD 123701    & 1.46       & 1.73       & 1.79       & 1.62      & 1.68       & 0.15       &         & 1 \\
             & 1.56       & 1.20       & -          & 1.14      & -          & -          &           & 2\\
HD 176021    & 1.33       & 1.47       & -          & 1.32      & 1.50       & 0.38       &           & 1\\
             & 0.48       & 0.52       & 0.87       & 0.62      & 0.53       & 0.14       &           & 6\\
HD 188985    & 1.39       & 1.56       & -          & 1.37      & -          & 0.10       &           & 1\\
             & 1.16       & 1.23       & 0.82       & 1.04      & 0.72       & 0.29       &           & 5 \\
             & -          & 0.93       & -          & 0.89      & -          & -          &           & 7\\ 
HD 202020    & 0.75       & 0.82       & -          & 0.71      & -          & 0.07       &           & 1\\ 
             & 0.67       & 0.03       & 0.29       & 0.25      & $-$0.04    &  -         &          & 8 \\       
\hline   
\end{tabular}
 
\textit{Note.} References: 1. Our work, 2. de Castro (2016), 3. Vanture(1992), 4. Pereira \& Junqueira (2003), 
5. Allen \& Barbuy (2006a), 6. Sneden \& Bond (1976), 7. North et al. (1994b), 8 Luck \& Bond (1991)\\
\end{table*}
}                    

\section{Conclusion}
Abundance analysis results of 12 potential CH star candidates,
11 from the CH stars catalogue of Bartkevicious (1996) and 
one object HE 0308$-$1612 from Goswami et al. (2010)  are presented. 
Detailed analysis shows two of them, CD$-$28 1082 and CD$-$38 2151 
to be very metal-poor   and highly enhanced in carbon. Five
objects HE~0308$-$1612, HD~30443, HD~87853, HD~176021 and HD~202020  
show characteristic properties of CH stars.  Estimated C/O ratios 
for these objects are similar to those  generally noticed in
 CH stars.  While HE~0308$-$1612 
and HD 30443 are CH giants,  HD 87853  and HD 176021 are
CH subgiants, and  HD 202020 is in
turn off stage.  The abundance patterns of these stars 
match  well with the abundance patterns of CH stars from literature.

While HE~0308$-$1612, 
and HD~176021  show moderate enhancement of carbon, HD~202020 shows 
near solar carbon abundance. Among the CH stars, HD~30443 shows the
 highest  carbon enhancement with [C/Fe] ${\sim}$ 1.68. 
The remaining five objects, HD 29370, HD 50264, HD 87080, HD 123701
and HD 188985 are found to show the characteristic properties
of barium stars. Among the barium stars, only one object HD 87080 
is found to be moderately metal-poor ([Fe/H] = $-$0.47), the rest four
are  near solar.  Barium  stars with [Ba/Fe] $>$ 0.60
are classified as  strong Ba stars (Yang et al. 2016). All the Ba stars 
in our sample show [Ba/Fe] $>$ 1 and are 
Ba strong. The abundance patterns of the Ba stars  match well
with the abundance patterns of strong Ba stars in the sample of 
de Castro et al. (2016) and Yang et al. (2016). de Castro et al. (2016)
shows that 90 percent of the Ba stars in their sample belong to thin disk 
population. From kinematic analysis, all the five  Ba stars are found to 
belong to thin disk population.  

We could estimate the carbon isotopic ratio  $^{12}$C/$^{13}$C 
only for six objects, two  CEMP, two CH and two barium stars. The
values are found to be in the range 7.4-16.
The stars for which we could estimate $^{12}$C/$^{13}$C are either 
in the giant or the subgiant phase of evolution. During the H burning 
stage of the main-sequence  stars, proton capture of a  $^{12}$C can 
take place  leading to the production of $^{13}$C. According to 
the standard evolutionary theory,
the convective envelope of the star, which is ascending the giant branch,
can expand inward and mix the CNO processed material from the burning
shell to the outer envelope. As a result of this first dredge up, the
carbon isotopic ratio on the surface of the star may decrease. This may
explain the low $^{12}$C/$^{13}$C values estimated for our programme stars.
The low values of $^{12}$C/$^{13}$C ($<$ 15) may also
be explained by certain nucleosynthesis  of the material
contributed by the donor star, in which the accreted material is
mixed into the hydrogen burning region either during the main-sequence
or the first ascent of the  giant branch (Vanture, 1992).

 Nitrogen is enhanced in all the stars other than HD~29370. 
Nitrogen production is believed to be mostly through CN cycling.
 First dredge up brings the CNO processed nitrogen to the surface as 
the star evolves to the red giant branch. Because of this, in normal 
giants the  nitrogen is found to be enhanced and  $^{12}$C/$^{13}$C  
reduced. 
A second distinct mixing episode  can happen 
when the star becomes brighter than the RGB bump (Gratton et al. 2000).
This can also reduce the observed $^{12}$C/$^{13}$C ratio to nearly 6-10
and the abundance of nitrogen can raise by nearly a factor of 4.
 The  abundance pattern is expected to be 
the same until the AGB phase of evolution where the second/third dredge-up 
can cause changes in the surface chemical compositions. Contrary to this,
observations show that the abundances of light elements such as C, N, 
O and the carbon isotopic ratio are modified at the RGB bump. 
This is exhibited by more 
than 95 percent of the low-mass stars and it is found to be independent of 
the stellar environment (Charbonnel \& do Nascimento, 1998). Zahn (1992) 
proposed that the interaction between meridional circulation and 
turbulence induced rotation can be the cause of mixing and the 
subsequent changes in the surface chemical compositions. Charbonnel (2005) 
suggested that this extra mixing may be rotation induced. But none of 
these processes alone could explain the observations. Based on the 3D 
modelling of low-mass RGB stars  (Dearborn, Lattanzio \& Eggleton 2006), Eggleton, Dearborn \& Lattanzio (2006) suggested that the molecular-weight inversion created by 
$^{3}$He($^{3}$He, 2p) $^{4}$He reaction in the outer regions of H 
burning shell can cause a transport mechanism often referred  as 
thermohaline mixing that may cause  a change in the surface chemical 
compositions of C, N, O, and Na. 
The effect of thermohaline mixing on the surface abundances of a
low-metallicity, low-mass star that has accreted material from a carbon rich
companion has also been  investigated  by Stancliffe  et al. (2007) based on 
binary models. It is commonly assumed that a  star that accreted 
material keep it on the surface without any mixing with the interior
until the first dredge up; however, thermohaline mixing can happen
that causes the mixing of accreted material with the rest of the star
as it has a higher mean molecular weight.

All the stars in our sample show enhancement in heavy elements. 
As their evolutionary status does not  support the observed 
enhanced abundances of heavy elements, the enhancement  may be attributed  
to an invisible binary companion. In our 
sample, only  HD~202020 is  a confirmed binary (McClure 1997).

The abundance results presented here will add to  the abundance data 
obtained from our previous studies in a homogeneous manner and 
compiled in Cristallo et al. (2016), that  
 can be used to constrain the physics and the nucleosynthesis
occuring in low-mass AGB stars. Abundance results of CEMP-r/s stars
will  provide observational  constrains for theoretical studies of
i-process,  suggested in the recent years for the origin of 
CEMP-r/s stars.

\section{ACKNOWLEDGEMENT}
 We thank the staff at IAO and at the remote control station at CREST,
Hosakote, and the staff at VBO, Kavalur, for assisting during the 
observations. Funding 
from DST SERB project No. EMR/2016/005283 is gratefully acknowledged.
We thank the anonymous referee for many constructive suggestions.
We would like to acknowledge Abha K for performing some initial 
calculations on a few stars while at IIA during 2015. 
This work made use of the SIMBAD astronomical database, operated
at CDS, Strasbourg, France,  the NASA ADS, USA and  
 data from the European Space Agency (ESA) 
mission Gaia (https://www.cosmos.esa.int/gaia), processed by the Gaia 
Data Processing and Analysis Consortium 
(DPAC, https://www.cosmos.esa.int/web/gaia/dpac/consortium). T.M. 
acknowledges support provided by the Spanish Ministry of Economy
and Competitiveness (MINECO) under grant AYA-2017-88254-P. Based on data collected using VBT echelle, HESP, and FEROS.
{}

\newpage

{\footnotesize
\begin{table*}
{\bf Table A1 : Elemental abundances in HE 0308$-$1612, CD$-28$ 1082 and HD 29370}
\resizebox{\textwidth}{!}{\begin{tabular}{|ccc|ccc|ccc|ccc|}
\hline
  &      &       & \multicolumn{3}{c}{HE 0308$-$1612} & \multicolumn{3}{c}{CD$-28$ 1082} & \multicolumn{3}{c}{HD 29370} \\
\hline
  & Z & Solar log$\epsilon^{a}$ & log$\epsilon$& [X/H]& [X/Fe]&log$\epsilon$& [X/H]    &[X/Fe]& log$\epsilon$  & [X/H] & [X/Fe] \\ \hline
C     & 6  & 8.43  & 8.48(syn)         & 0.05    & 0.78 & 8.16(syn)         & $-0.27$      & 2.19 & 8.37(syn)        & $-0.06$      & 0.09\\ 
N     & 7  & 7.83  & 7.25(syn)         & $-$0.58 & 0.15 & 8.10(syn)         & 0.27         & 2.73 & 7.70(syn)        & $-0.13$      & 0.02 \\
O     & 8  & 8.69  & -                 & -       &   -  & -                 & -            &   -  &  -        &   -          & -   \\              
Na I	& 11 &	6.24 & 5.85(syn)         & $-$0.39 & 0.34 & 5.59 (syn)        & $-2.01$      & 0.45 & 6.32(1)          & 0.08         & 0.23\\
Mg I	& 12 &	7.60 & 7.22$\pm$ 0.01(2) & $-$0.38 & 0.35 & 5.59 (syn)        & $-2.01$      & 0.45 & 7.47(1)          & $-0.13$      & 0.02\\
Si I	& 14 &	7.51 &   -               &   -     &   -  & -                 & -            &   -  & 7.00$\pm$0.04(2) & $-0.51$      & $-0.36$\\
Ca I	& 20 &	6.34 &5.98$\pm$ 0.18(5)  & $-$0.36 & 0.37 & 4.15$\pm$ 0.12(4) & $-2.19$      & 0.27 & 6.00$\pm$0.02(3) & $-0.34$      & $-0.19$\\
Sc II	& 21 &	3.15 &2.95(1)            & $-$0.2  & 0.53 & -                 & -            &   -  & 3.21$\pm$0.16(3) & 0.06         & 0.19\\
Ti I	& 22 &	4.95 &4.54$\pm$ 0.15(5)  & $-$0.41 & 0.32 & -                 & -            &   -  & 4.52$\pm$0.05(4) & $-0.43$      & $-0.28$\\
Ti II	& 22 &	4.95 &4.52(1)            & $-$0.43 & 0.30 & -                 & -            &   -  & 4.54$\pm$0.14(2) & $-0.41$      & $-0.28$\\
V I	  & 23 &	3.93 &3.49(syn)          & $-$0.44 & 0.29 & -                 & -            &   -  & 4.07$\pm$0.14(3) & 0.14         & 0.29   \\
Cr I	& 24 &	5.64 &4.53$\pm$ 0.20(3)  & $-$1.11 &$-$0.39& -                & -            &   -  & 5.54$\pm$0.09(3) & $-0.10$      & 0.05 \\
Cr II	& 24 &	5.64 & -                 &         &      & -                 & -            &   -  & 5.33(1)          & $-0.31$      & $-0.18$ \\
Mn I	& 25 &	5.43 &4.99$\pm$ 0.13(3)  & $-$0.44 & 0.29 &  4.45(syn)        & $-0.98$      & 1.48 & 5.60(syn)        & $0.17$       & $0.32$ \\
Fe I	& 26 &	7.50 &6.78$\pm$ 0.19(40) & $-$0.72 & -    & 5.04$\pm$ 0.08(13)& $-2.46$      & -    & 7.35$\pm$0.11(35)& $-0.15$      & - \\
Fe II	& 26 &	7.50 &6.77$\pm$ 0.15(3)  & $-$0.73 & -    & 5.06$\pm$ 0.02(2) & $-2.44$      & -    & 7.37$\pm$0.04(5) & $-0.13$      & -\\
Co I	& 27 &	4.99 & -                 &  -      & -    & -                 & -            &   -  & 4.83$\pm$0.09(2) & $-0.16$      & $-0.01$ \\
Ni I	& 28 &	6.22 &5.81$\pm$ 0.13(5)  & $-$0.41 & 0.32 & -                 & -            &   -  & 6.01$\pm$0.13(4) & $-0.21$      & $-0.06$ \\
Zn I	& 30 &	4.56 &3.65(syn)          &$-$0.91  &$-$0.19&-                 & -            &   -  & 4.51(1)          & $-0.05$      & 0.10 \\
Sr I	& 38 &	2.87 &3.24(1)            &0.37     & 1.1  & 1.85(syn)         & $-1.02$      & 1.44 & 3.54(syn)        & 0.67         & 0.82 \\
Y II	& 39 &	2.21 &2.61$\pm$0.09(4)   & 0.4     & 1.13 & 1.38(1)           & $-0.83$      & 1.61 & 2.95$\pm$0.14(2) & 0.74         & 0.87 \\
Zr II	& 40 &	2.58 & -                 &  -      &   -  & -                 & -            &   -  & 4.09(1)          & 1.51         & 1.64 \\
Ba II & 56 & 2.18  & 3.08(syn)         & 0.9     & 1.63 & 1.83(syn)         & $-0.35$      & 2.09 & 3.58(syn)        & 1.40         & 1.53 \\
La II & 57 & 1.10  & 1.80(syn)         & 0.7     & 1.43 & 0.21(syn)         & $-0.89$      & 1.55 & 2.10(syn)        & 1.00         & 1.13 \\
Ce II	& 58 &	1.58 &2.56$\pm$0.08(3)   & 0.98    & 1.71 & 1.11$\pm$ 0.07(2) & $-0.47$      & 1.97 & 2.59$\pm$0.11(2) & 1.01         & 1.14 \\
Pr II & 59 & 0.72  & 2.00(syn)         & 1.28    & 2.01 & 0.58(1)           & $-0.14$      & 2.30 &  -               &  -           & -\\
Nd II & 60 & 1.42  & 2.42$\pm$0.18(10) & 1.0     & 1.73 & 0.97$\pm$ 0.12(7) & $-0.45$      & 1.99 & 2.48$\pm$ 0.05(3)& 1.06         & 1.19 \\
Sm II & 62 & 0.96  & 1.68$\pm$ 0.14(2) & 0.72    & 1.45 & 0.81(1)           & $-0.15$      & 2.29 & 2.24(1)          & 1.28         & 1.41 \\
Eu II & 63 & 0.52  & 0.14(syn)         &$-$0.38  & 0.35 & 0.15(syn)         & $-0.37$      & 2.07 & -                & -            & - \\

\hline
\end{tabular}
}

{\bf Table A2 : Elemental abundances in HD 30443, CD$-38$ 2151 and HD 50264 }
\resizebox{\textwidth}{!}{\begin{tabular}{|ccc|ccc|ccc|ccc|}
\hline
  &      &       & \multicolumn{3}{c}{HD 30443} & \multicolumn{3}{c}{CD$-38$ 2151} & \multicolumn{3}{c}{HD 50264} \\
\hline
  & Z & Solar log$\epsilon^{a}$ & log$\epsilon$& [X/H]& [X/Fe]&log$\epsilon$& [X/H]    &[X/Fe]& log$\epsilon$  & [X/H] & [X/Fe] \\ \hline
C     & 6  & 8.43  & 8.43(syn)         & 0.00    & 1.68 & 7.9(syn)          & $-0.53$ & 1.50 & 8.50(syn)         & 0.07    & 0.21\\
N     & 7  & 7.83  & 6.55(syn)         & $-$1.28 & 0.40 & 7.2(syn)          & $-0.63$ & 1.40 & 8.35(syn)         & 0.52    & 0.66 \\
O     & 8  & 8.69  & 8.42(syn)         & $-0.27$ & 1.41 & 7.22(syn)         & $-1.47$ & 0.56 & 8.89(syn)         & 0.20    & 0.34 \\
Na I	& 11 &	6.24 & 5.14$\pm$ 0.20(3) & $-$1.1  & 0.58 & -                 &    -    &  -   & -                 &    -    &  -   \\
Mg I	& 12 &	7.60 & 6.04$\pm$ 0.13(2) & $-$1.56 & 0.12 & 6.20(syn)         & $-1.40 $ & 0.63& 7.60$\pm$ 0.10(3) & 0.00    & 0.14\\
Si I	& 14 &	7.51 & 6.65$\pm$ 0.05(2) & $-$0.86 & 0.82 & 7.10(syn)         & $-0.41$ & 1.62 & -                 &    -    &  -   \\ 
Ca I	& 20 &	6.34 & 5.17$\pm$ 0.17(4) & $-$1.17 & 0.51 & 4.89$\pm$ 0.08(2) & $-1.45$ & 0.58 & 6.08$\pm$ 0.12(8) & $-0.26$ & $-0.12$\\
Sc II	& 21 &	3.15 & 1.60(syn)         & $-$1.55 & 0.13 & 1.30$\pm$ 0.12(2) & $-1.86$ & 0.18 & 2.88(syn)         & $-0.27$ & $-0.16$ \\
Ti I	& 22 &	4.95 & 3.37$\pm$ 0.16(9) & $-$1.58 & 0.10 & 3.63$\pm$ 0.06(3) & $-1.32$ & 0.71 & 4.85$\pm$ 0.06(6) & $-0.10$ & 0.04\\
Ti II	& 22 &	4.95 & 3.68$\pm$ 0.25(4) & $-$1.27 & 0.41 & 3.21$\pm$ 0.17(2) & $-1.74$ & 0.29 & 4.80$\pm$ 0.09(6) & $-0.15$ & $-0.04$\\
V I	  & 23 &	3.93 & 1.61(syn)         & $-$2.32 &$-$0.64&3.06$\pm$ 0.07(2) & $-0.87$ & 1.16 & -                 &    -    &  -   \\
Cr I	& 24 &	5.64 & -                 &   -     & -    & 3.91$\pm$ 0.06(5) & $-1.73$ & 0.30 & 5.54$\pm$ 0.10(9) & $-0.10$ & 0.04 \\
Cr II	& 24 &	5.64 & -                 &   -     & -    &  -                &  -      &  -   & 5.56$\pm$ 0.06(3) & $-0.08$ & 0.03 \\
Mn I	& 25 &	5.43 & 3.88(syn)         & $-$1.61 & 0.07 & 3.20(syn)         & $-2.23$ &$-0.20$&4.99(syn)         & $-0.44$ & $-0.30$\\
Fe I	& 26 &	7.50 & 5.82$\pm$ 0.05(11)& $-$1.68 & -    & 5.47$\pm$ 0.10(12)& $-2.03$ & -    & 7.36$\pm$ 0.09(93)& $-0.14$ & - \\
Fe II	& 26 &	7.50 & 5.81$\pm$ 0.11(3) & $-$1.69 & -    & 5.47(1)           & $-2.03$ & -    & 7.39$\pm$ 0.07(11)& $-0.11$ & - \\
Co I	& 27 &	4.99 & 3.45(syn)         & $-$1.54 & 0.14 & 3.11(1)           & $-1.88$ & 0.15 & 4.76 $\pm$ 0.02(2)& $-0.23$ & $-0.09$ \\
Ni I	& 28 &	6.22 & 5.13$\pm$ 0.22(3) & $-$1.09 & 0.59 & 4.06(1)           & $-2.16$ & $-0.13$&6.01$\pm$ 0.17(7)& $-0.21$ & $-0.07$ \\
Zn I	& 30 &	4.56 & 3.22 (1)          &$-$1.34  & 0.34 &  -                &  -      &  -   & 4.67$\pm$0.06 (2) & 0.11    & 0.25 \\ 
Sr I	& 38 &	2.87 & 1.93(syn)         &$-$0.94  & 0.74 & 2.00(syn)         & $-0.87$ & 1.16 & 4.63(syn)         & 1.76    & 1.90 \\
Y II	& 39 &	2.21 & 1.90$\pm$0.04(2)  & $-$0.31 & 1.37 & 0.75$\pm$ 0.14(2) & $-1.46$ & 0.57 & 3.26 $\pm$ 0.08(7)& 1.05    & 1.16 \\
Zr II	& 40 &	2.58 & 2.50(syn)         & $-$0.08 & 1.60 & 2.55(1)           & $-0.03$ & 2.00 &   -               & -    &  -      \\           
Ba II	& 56 &	2.18 & 2.20(syn)         & 0.02    & 1.70 & 1.12(syn)         & $-1.06$ & 0.97 & 3.47(syn)         & 1.29    & 1.40 \\
La II	& 57 &	1.10 & -                 &   -     & -    & $-0.05$(syn)      & $-1.15$ & 0.88 & 2.29(syn)         & 1.19    & 1.30 \\
Ce II	& 58 &	1.58 & 1.52$\pm$0.12(5)  & $-0.06$ & 1.62 & 0.79$\pm$ 0.07(2) & $-0.79$ & 1.24 & 3.13 $\pm$ 0.05(2)& 1.55    & 1.66 \\
Pr II & 59 & 0.72  & 1.71$\pm$ 0.07(4) & 0.98    & 2.67 & 0.21(1)           & $-0.51$ & 1.52 &    -              & -       &  -    \\            
Nd II	& 60 &	1.42 & 2.21$\pm$0.18(9)  & 0.79    & 2.47 & 0.73$\pm$ 0.17(3) & $-0.69$ & 1.34 & 2.66 $\pm$ 0.09(4)& 1.24    & 1.35 \\
Sm II	& 62 &	0.96 & 1.52$\pm$ 0.17(5) & 0.56    & 2.24 & 0.46(1)           & $-0.50$ & 1.53 &    -              & -       &  -    \\            
Eu II & 63 &  0.52 & -                 &   -     & -    &  -                &  -      &  -   &     -             & -       &  -     \\           
\hline
\end{tabular}
}

$^a$  Asplund (2009), The number inside the  parenthesis shows the 
number of lines used for the abundance determination. 
\end{table*}
}

{\footnotesize
\begin{table*}
  {\bf Table A3 : Elemental abundances in HD 87080, HD 87853 and HD 123701.}
 \resizebox{\textwidth}{!}{\begin{tabular}{|ccc|ccc|ccc|ccc|}
\hline
  &      &       & \multicolumn{3}{c}{HD 87080} & \multicolumn{3}{c}{HD 87853} & \multicolumn{3}{c}{HD 123701} \\
\hline
    & Z & Solar log$\epsilon^{a}$ & log$\epsilon$& [X/H]& [X/Fe]&log$\epsilon$& [X/H]    &[X/Fe]& log$\epsilon$  & [X/H] & [X/Fe] \\ \hline
C     & 6  & 8.43  & 8.05(syn)         & $-0.38$ & 0.09 &  -                  &  -      &  -   & 8.53(syn) & 0.10 & 0.05 \\
N     & 7  & 7.83  & 8.25(syn)         & 0.42    & 0.89 &  -                  &  -      &  -   &  8.75(syn) & 0.92 & 0.87 \\
O     & 8  & 8.69  & 8.59(syn)         & $-0.10$ & 0.37 & 8.35(syn)           & $-$0.34 & 0.39 & 9.36(syn)         & 0.67 & 0.62 \\    
Na I	& 11 &	6.24 & 5.85$\pm$ 0.10(2) & $-0.39$ & 0.08 & 5.92(syn)           & $-$0.32 & 0.41 & 6.37$\pm$ 0.01(2) & 0.13 & $0.08$  \\
Mg I	& 12 &	7.60 & 7.30$\pm$ 0.06(2) & $-0.30$ & 0.17 & 7.55(1)             & $-$0.05 & 0.68 & 7.64(1)              & 0.04 & $-0.01$\\
Si I	& 14 &	7.51 & 6.60$\pm$ 0.11(3)& $-0.91$  &$-0.44$& -                  &  -      &  -   &   -                  &  -      &  - \\
Ca I	& 20 &	6.34 & 5.79$\pm$ 0.11(8)& $-0.55$  &$-0.08$& 5.90$\pm$ 0.20(12) & $-$0.44 & 0.29 & 6.14$\pm$ 0.05(4) & $-0.20$ & $-0.25$\\
Sc II	& 21 &	3.15 & 2.63(syn)        & $-0.52$  &$-0.03$& 2.03(syn)          & $-$1.15 &$-$0.39& 3.25(syn)             & 0.10    & 0.03 \\
Ti I	& 22 &	4.95 & 4.53$\pm$ 0.13(5)& $-0.42$  & 0.05 & 4.34(1)             & $-$0.61 & 0.12 & 5.14$\pm$ 0.18(7) & 0.19    & 0.14 \\
Ti II	& 22 &	4.95 & 4.48$\pm$ 0.09(4)& $-0.47$  & 0.02 & 4.33$\pm$ 0.12(5)   & $-$0.62 & 0.11 & 5.34$\pm$ 0.20(3) & 0.39    & 0.32 \\
V I	  & 23 &	3.93 & -                &  -       &  -   & 3.24(1)             & $-$0.69 & 0.04 & 4.10$\pm$ 0.06(3) & 0.17 & 0.12\\                
Cr I	& 24 &	5.64 & 5.14$\pm$ 0.08(3)& $-0.50$  &$-0.03$& 4.81$\pm$ 0.05(5)  & $-$0.83 &$-$0.10& 5.76$\pm$ 0.09(3) & 0.12    & 0.07 \\
Cr II	& 24 &	5.64 &5.07$\pm$ 0.04(3) & $-0.57$  &$-0.08$& 4.77(1)            & $-$0.87 &$-$0.14&    -                  &  -      &  - \\
Mn I	& 25 &	5.43 &4.69(syn)         & $-0.74$  &$-0.27$& 4.33(syn)          & $-$1.10 &$-$0.37& 5.25(syn)         & $-0.18$ & $-0.23$\\
Fe I	& 26 &	7.50 &7.03$\pm$ 0.09(65)& $-0.47$  & -    & 6.78$\pm$ 0.09(53)  & $-$0.72 & -    & 7.55$\pm$ 0.09(38)& 0.05    & -\\
Fe II	& 26 &	7.50 &7.00$\pm$ 0.04(10)& $-0.49$  & -    & 6.76$\pm$ 0.09(2)   & $-$0.74 & -    & 7.57$\pm$ 0.09(13)& 0.07    & -\\
Co I	& 27 &	4.99 & 4.41 $\pm$ 0.09(2& $-0.58$  &$-0.11$& 4.30(1)            & $-$0.69 & 0.04 & 5.19(1)           & 0.20    & 0.15 \\
Ni I	& 28 &	6.22 & 5.74$\pm$ 0.18(6)& $-0.48$  &$-0.01$& 5.45$\pm$ 0.13(8)  & $-$0.77 &$-$0.04& 6.22$\pm$ 0.19(6)& $0.0$ & $-0.05$\\
Zn I	& 30 &	4.56 & 4.10(1)          & $-0.46$  &0.01  & 3.75$\pm$0.02 (2)   &$-$0.81  &$-$0.08& 4.50(1)          & $-0.06$ & $-0.11$ \\
Sr I	& 38 &	2.87 & 4.24(syn)        & 1.37     &1.84  &  -                  &  -      &  -   & 4.50 (syn)          & 1.63    & 1.58  \\
Y II	& 39 &	2.21 & 2.84$\pm$ 0.07(7)& 0.63     & 1.12 & 1.00$\pm$0.05(3)    & $-$1.21 &$-$0.48& 3.45$\pm$ 0.14(3) & 1.24    & 1.17\\
Zr II	& 40 &	2.58 & -                &  -       &  -   &   -                  &  -      &  -   & 3.48(1)           & 0.90    & 0.83 \\
Ba II	& 56 &	2.18 & 3.30(syn)        & 1.12     & 1.61 & 1.02(syn)           &$-$1.16  &$-$0.43& 3.87(syn)           & 1.69    & 1.62\\
La II	& 57 &	1.10 & 2.36(syn)        & 1.26     & 1.75 &   -                  &  -      &  -   & 2.63(syn)          & 1.53    & 1.46 \\
Ce II	& 58 &	1.58 & 2.80$\pm$ 0.04(4)& 1.22     & 1.71 &   -                  &  -      &  -   & 3.38$\pm$ 0.04(4) & 1.80    & 1.73 \\
Pr II & 59 & 0.72  & -                &  -       &  -   &   -                  &  -      &  -   & 2.58(1)           & 1.86    & 1.79 \\
Nd II	& 60 &	1.42 &2.40$\pm$ 0.15(10)& 0.98     & 1.47 & 1.63(1)             & 0.21    & 0.94  & 3.11$\pm$ 0.15(5) & 1.69    & 1.62\\
Sm II	& 62 &	0.96 & 2.15$\pm$ 0.17(3)& 1.19     & 1.68 &   -                  &  -      &  -   & 2.71(1)           & 1.75    & 1.68\\ 
Eu II & 63 & 0.52  & 0.39(syn)        & $-0.13$  & 0.36 & 0.13(1)             & $-$0.39 & 0.34  & 0.74(syn)         & 0.22    & 0.15 \\
 \hline
\end{tabular}
}

{\bf Table A4 : Elemental abundances in HD 176021, HD 188985 and HD 202020}
 \resizebox{\textwidth}{!}{\begin{tabular}{|ccc|ccc|ccc|ccc|}
\hline
  &      &       & \multicolumn{3}{c}{HD 176021} & \multicolumn{3}{c}{HD 188985} & \multicolumn{3}{c}{HD 202020} \\
\hline
    & Z & Solar log$\epsilon^{a}$ & log$\epsilon$& [X/H]& [X/Fe]&log$\epsilon$& [X/H]    &[X/Fe]& log$\epsilon$  & [X/H] & [X/Fe] \\ \hline
C     & 6  & 8.43  & 8.33(syn)         & $-0.10$ & 0.52 & 8.61(syn)         & 0.18    & 0.17  & 8.70(syn) & 0.27 & 0.23 \\
N     & 7  & 7.83  & 7.80(syn)         & $-0.03$ & 0.59 & 8.55(syn)         & 0.72    & 0.71  & 8.44(syn) & 0.61 & 0.57 \\
O     & 8  & 8.69  &  -                &    -    &  -   & 8.66(syn)         & $-0.03$ &$-0.04$& 8.40(syn)        & $-0.29$ & $-0.33$ \\ 
Na I	& 11 &	6.24 & 5.80$\pm$ 0.01(2) & $-0.44$ & 0.18 &   -               & -       &  -    & 6.35$\pm$ 0.06(3) & 0.11 & 0.07 \\
Mg I	& 12 &	7.60 & 7.36$\pm$ 0.06(3) & $-0.24$ & 0.38 & 7.67$\pm$ 0.15(3) & 0.07    & 0.06  & 7.91$\pm$ 0.06(3) & 0.31 & 0.27\\
Si I	& 14 &	7.51 &  -                &    -    &  -   &   -               & -       &  -    &   -               & -       &  -  \\   
Ca I	& 20 &	6.34 & 5.85$\pm$ 0.14(6) & $-0.49$ & 0.13 & 6.15$\pm$ 0.10(10)& $-0.19$ &$-0.20$& 6.16$\pm$ 0.11(14) & $-0.18$ & $-0.22$\\
Sc II	& 21 &	3.15 & 2.54(syn)         & $-0.61$ & 0.01 & 2.95(syn)         & $-0.20$ &$-0.21$& 3.14(syn)           & $-0.01$ & $-0.05$ \\
Ti I	& 22 &	4.95 & 4.62$\pm$ 0.09(3) & $-0.33$ & 0.29 & 4.90$\pm$ 0.10(4) & $-0.05$ &$-0.06$&  5.11$\pm$ 0.09(7) & 0.16 & 0.12 \\
Ti II	& 22 &	4.95 & 4.72$\pm$ 0.12(2) & $-0.23$ & 0.42 & 4.83$\pm$ 0.14(3) & $-0.12$ &$-0.13$& 5.05$\pm$ 0.09(6) & 0.10 & 0.02\\
V I	  & 23 &	3.93 & 2.88(syn)         & $-1.05$ &$-0.43$&   -              & -       &  -    &  3.79(syn)              & $-0.14$ & $-0.18$\\
Cr I	& 24 &	5.64 & 5.06$\pm$ 0.09(3) & $-0.58$ & 0.04 & 5.82$\pm$ 0.06(6) & 0.18    & 0.17  & 5.71$\pm$ 0.09(7) & 0.13 & 0.09 \\
Cr II	& 24 &	5.64 & 4.93$\pm$ 0.08(2) & $-0.71$ & 0.06 &   -               & -       &  -    &   -               & -       &  -  \\
Mn I	& 25 &	5.43 & 4.98(syn)         & $-0.45$ & 0.17 & 5.06(syn)         & $-0.37$ &$-0.38$& 5.15(syn)           & $-0.28$ & $-0.32$\\
Fe I	& 26 &	7.50 & 6.88$\pm$ 0.08(37)& $-0.62$ & -    & 7.51$\pm$ 0.10(45)& $0.01$  & -     & 7.54$\pm$ 0.09(120) & $0.04$ & -\\
Fe II	& 26 &	7.50 & 6.84$\pm$ 0.05(2) & $-0.65$ & -    & 7.51$\pm$ 0.09(16)& $0.01$  & -     & 7.58$\pm$ 0.03(4) & $0.08$ & -\\
Co I	& 27 &	4.99 & 4.66(1)           & $-0.33$ & 0.29 & 5.33(1)           & 0.34    & 0.33  & 5.00$\pm$ 0.07(3) & 0.01 & $-0.03$\\
Ni I	& 28 &	6.22 & 5.64$\pm$ 0.15(4) & $-0.58$ & 0.04 & 6.33$\pm$ 0.19(6) & 0.11    & 0.10  & 6.27$\pm$ 0.08(7) & 0.05 & 0.01\\
Zn I	& 30 &	4.56 & 4.23$\pm$0.09(2)  & $-0.33$ & 0.29 & 4.82$\pm$0.11(2)  &0.22     & 0.21  & 5.00$\pm$0.13(2)           &0.44 & 0.40 \\
Sr I	& 38 &	2.87 & 4.14(syn)         & 1.27    & 1.89 & 4.95(syn)         & 2.08    & 2.07  & 4.73(syn)               & 1.86 & 1.82\\
Y II	& 39 &	2.21 & 2.67$\pm$ 0.11(3) & 0.46    & 1.11 & 3.35$\pm$0.12(6)  & 1.14    & 1.13  & 3.25$\pm$0.05(6) & 1.04 & 1.00 \\
Zr II	& 40 &	2.58 &  -                &    -    &  -   &   -               & -       &  -    & 3.14(1)     & 0.56 & 0.48 \\
Ba II	& 56 &	2.18 & 2.88(syn)         & 0.70    & 1.35 & 3.65(syn)         & 1.47    & 1.46  & 3.06(syn)        & 0.88 & $0.80$ \\
La II	& 57 &	1.10 & 1.78(syn)         & 0.68    & 1.33 & 2.50(syn)         & 1.40    & 1.39  & 1.93(syn)         & 0.83 & 0.75\\
Ce II	& 58 &	1.58 & 2.40$\pm$ 0.17(3) & 0.82    & 1.47 & 3.15$\pm$0.21(5)  & 1.57    & 1.56  & 2.48$\pm$0.06(2) & 0.90 & 0.82\\
Pr II & 59 &  0.72 & -                 &  -      & -    &   -               & -       &  -    &   -               & -       &  -  \\ 
Nd II	& 60 &	1.42 & 2.09$\pm$ 0.04(4) & 0.67    & 1.32 & 2.80$\pm$0.06(3)  & 1.38    & 1.37  & 2.21$\pm$0.18(5) & 0.79 & 0.71 \\
Sm II	& 62 &	0.96 & 1.81$\pm$ 0.04(2) & 0.85    & 1.50 &   -               & -       &  -    &   -               & -       &  -  \\
Eu II & 63 & 0.52  & 0.25(syn)         & $-0.27$ & 0.38 & 0.63(syn)         & 0.11    & 0.10  & 0.67(syn)    & 0.15 & 0.07 \\
\hline
\end{tabular}
}

$^a$  Asplund (2009), The number inside the  parenthesis shows 
the number  of lines used for the abundance determination. 
\end{table*}
}

{\footnotesize
\begin{table*}
{\bf Table A5: Equivalent widths (in m\r{A}) of Fe lines used for deriving 
atmospheric parameters.}\\ 
\resizebox{\textwidth}{!}{\begin{tabular}{cccccccccccccc}
\hline                       
Wavelength(\r{A}) & Element & $E_{low}$(eV) & log gf & CD-28 1082&HD 29370 & CD-38 2151 & HD 50264 & HD 87080 & HD 123701 & HD 176021 & HD 188985 & HD 202020 & References  \\ 
\hline 
4114.446 & Fe I & 2.831 & $-1.220$& - &- & - & - & - & - & - & 72.9(7.34) & - & 1 \\
4139.927 & & 0.990 & $-3.629$&- &- & - & - & - & - & 43.5(6.81) & - & 67.4(7.58) &1\\ 
4184.891 & & 2.831 & $-0.860$&- & - & - & - & - & - & - & 99.0(7.50) &- & 1\\
4348.937 & & 2.990 & $-2.130$&- & - & - & 41.0(7.20)& - & - & - & 35.6(7.31) &48.7(7.39)& 1 \\ 
4566.514 & & 3.301 & $-2.250$&- & - & - & 28.2(7.27)& - & - & - & - & 37.9(7.52)& 1\\  
\hline
\end{tabular}}

The numbers in the  parenthesis in columns 5-13 give the derived 
abundances from the respective lines. \\
References: 1. F{\"u}hr et al. (1988) \\
{\bf Note:} This table is available in its entirety in online only. A portion is shown here for guidance regarding its form and content.
\end{table*}
}

{\footnotesize

\begin{table*}

{\bf Table A6: Equivalent widths (in m\r{A}) of Fe lines used for deriving atmospheric parameters.}\\ 

\begin{tabular}{cccccccc}
\hline                       
Wavelength(\r{A}) & Element & $E_{low}$(eV) & log gf & HD 30443 & HE 0308-1612  & HD 87853 & References  \\ 
\hline 
4114.446 & Fe I & 2.831 & $-1.220$& - &- &- & 1 \\
4139.927 & & 0.990 & $-3.629$&- &- & -  &1\\ 
4147.669 & & 1.485 & $-2.104$&- &- &83.6(6.82)& 1\\
4184.891 & & 2.831 & $-0.860$&- & - & 83.8(6.79) & 1\\
4187.039 & & 2.449 & $-0.548$&- &- &108.4(6.78)& 1\\
\hline
\end{tabular}

The numbers in the  parenthesis in columns 5-7 give the derived abundances from the respective line. \\
References: 1. F\"uhr et al. (1988) \\
{\bf Note:} This table is available in its entirety in online only. A portion is shown here for guidance regarding its form and content.
\end{table*}
}

{\footnotesize

\begin{table*}

{\bf Table A7: Lines used for deriving elemental abundances}\\ 

\resizebox{\textwidth}{!}{\begin{tabular}{cccccccccccccc}

\hline                       

Wavelength(\AA) & Element & $E_{low}$(eV) & log gf & CD$-28$ 1082& HD 29370 &CD$-38$ 2151 & HD 50264 & HD 87080 & HD 123701 & HD 176021 & HD 188985 & HD 202020 & Reference \\ 

\hline 

5682.633 & Na I& 2.102& $-0.700$ &-          & 129.7(6.32)& -         & -         &70.2(5.92) &-            &54.5(5.81)  &-           & 93.2(6.41)&1\\

5889.951 &     & 0.000& 0.100    &166.7(4.56)& -          & -         &-          &-          &-            &-           &-           & -         &1\\

6154.226 &     & 2.102& $-1.560$ & -         &-           &-          &-          &-          &55.6(6.36)   &-           &-          &35.2(6.35)&1\\
6160.747 &     & 2.104  & $-1.260$& -        &-           &-          &-          &30.2(5.79) &76.8(6.38)   &24.2(5.79)  &-          &48.4(6.28)&1\\

4571.096 & Mg I& 0.000  & $-5.691$& 56.6(5.89)&-          &170.4(6.64)&92.9(7.60) &-          &-            &86.1(7.42)  &-          & 108(7.95)&2\\

4730.029 &     & 4.346  & $-2.523$ & -        &-          &-          &52.5(7.68) & -         &-            &-           &50.3(7.83) & 66.(7.95)&3\\

\hline

\end{tabular}}

The numbers in the  parenthesis in columns 5-13 give the derived 
abundances from the respective lines. \\

References:  1. Kurucz et al. (1975), 2. Laughlin et al. (1974), 3. Lincke et al. (1971)\\
{\bf Note:} This table is available in its entirety in online only. A portion is shown here for guidance regarding its form and content.
\end{table*}

}

{\footnotesize
\begin{table*}
{\bf Table A8 : Lines used for deriving elemental abundances}\\ 
\resizebox{\textwidth}{!}{\begin{tabular}{cccccccc}
\hline                       
Wavelength(\r{A}) & Element & $E_{low}$(eV) & log gf & HD 30443 & HE 0308-1612  & HD 87853 & References  \\ 
\hline 
5682.633 & Na I& 2.102& $-0.700$ &99.20(4.96)         & 108.9(6.00)& -        &1\\
5688.205 &     & 2.100& $-0.450$  & 135.5(5.11) & 115.5(6.82) & -  &  1\\
5889.951 &     & 0.000& 0.100    &-& -          & -          &1\\
6154.226 &     & 2.102& $-1.560$ & -         &-           &-         &1\\
6160.747 &     & 2.104  & $-1.260$& 86.60(5.36)        &-           &-          &1\\
4571.096 & Mg I& 0.000  & $-5.691$& -                  &-          &    -       &2\\
4702.991 & Mg I & 4.346  & $-0.666$& -                  & 177.7(7.22) & 142.2(7.55) & 3\\
\hline
\end{tabular}}
		
The numbers in the  parenthesis in columns 5-7 give the derived abundances from the respective line. \\
References:  1. Kurucz et al. (1975), 2. Laughlin et al. (1974), 3. Lincke et al. (1971)\\
{\bf Note:} This table is available in its entirety in online only. A portion is shown here for guidance regarding its form and content.
\\
\end{table*}
}

\end{document}